\documentclass[10pt]{article}
\usepackage[letterpaper]{geometry}

\usepackage{graphicx}
\usepackage{pdfpages}			% Inclusão de (páginas de) arquivos PDF no documento
\usepackage{lipsum}				% para geração de dummy text
\usepackage{amsfonts,amsmath,amssymb,latexsym}
\usepackage{mathrsfs}
\usepackage{enumerate}
\usepackage{mathtools}
\usepackage[normalem]{ulem}

\usepackage{url}

\usepackage{xcolor}
\definecolor{OliveGreen}{RGB}{0,102,0}
\usepackage{colortbl}

\usepackage{longtable}

\usepackage{algorithm}
\usepackage{algpseudocode} %[noend]
\usepackage{algorithmicx}


\algnewcommand{\Input}[1]{\Statex \textbf{input: } #1 }
\algnewcommand{\Output}[1]{\Statex \textbf{output: } #1 \Statex }
\algnewcommand{\return}[1]{\State\textbf{return} #1}


\usepackage{tikz}
\usetikzlibrary{shapes,shapes.multipart, calc,matrix,arrows,positioning,automata,snakes}
\usepackage{subfig}

\newcommand{\yI}{\mathcal{I}}  % name  I
\newcommand{\yS}{\mathcal{S}}  % name S
\newcommand{\yA}{\mathcal{A}}  
\newcommand{\yQ}{\mathcal{Q}}  
\newcommand{\yR}{\mathcal{R}} 
\newcommand{\yB}{\mathcal{B}}  
\newcommand{\yC}{\mathcal{C}}

\newcommand{\yT}{\mathcal{T}}

\newcommand{\yfail}{fail}

\tikzset{
	>=stealth',
	punkt/.style={
		circle,
		rounded corners,
		draw=black, thick,
		text width=1.5em,
		minimum height=2em,
		text centered},
	punkts/.style={
		circle,
		rounded corners,
		draw=black, thick, 
		text width=1em,
		minimum height=1em,
		text centered},
	invisible/.style={
		draw=none,
		text width=1.5em,
		minimum height=0em,
		text centered},
	inv/.style={
		draw=none,
		text width=2.5em,
		minimum height=3em,
		text centered},
	pil/.style={
		->,
		thick,
		shorten <=2pt,
		shorten >=2pt,}
}

\newcommand{\everest}{\text{\textsc{Everest}\,}}
\newcommand{\ioco}{\text{\bf ioco}\,\,}
\newtheorem{definition}{Definition}
\newtheorem{prop}{Proposition}
\newtheorem{lemma}{Lemma}
\newtheorem{theorem}{Theorem}

\usepackage{fullpage}

\usepackage{color}

\begin{document}

% !TeX spellcheck=en_US

\title{A Model-Based Testing Tool for Asynchronous Reactive  Systems\thanks{Supported by CAPES.}}

\author{Adilson Luiz Bonifacio\thanks{Computing Department, University of Londrina, Londrina, Brazil.} \and
	Camila Sonoda Gomes\thanks{Computing Department, University of Londrina, Londrina, Brazil.}}

\date{} % An optional date to appear under the author(s)

\maketitle

% !TeX spellcheck=en_US
\begin{abstract}
Reactive systems are characterized by the interaction with the environment, where the exchange of  the input and output stimuli, usually, occurs asynchronously. 
Systems of this nature, in general, require a rigorous testing activity 
over their developing process. 
Therefore model-based testing has been successfully applied over asynchronous reactive systems using Input Output Labeled Transition Systems (IOLTSs) as the basis. 
In this work we present a reactive testing tool to check conformance, generate test suites and run test cases using IOLTS models.  
Our tool can check whether the behavior of an implementation under test (IUT) complies with the behavior of its respective specification. 
We have implemented a classical conformance relation \ioco and a more general notion of conformance based on regular languages. 
Further, the tool provides a test suite generation in a black-box testing setting for finding faults over IUTs according to a specific domain. 
We have also described some case studies to probe the tool's functionalities and also to highlight a comparative analysis on both conformance approaches.
Finally, we offer experiments to evaluate the performance of our tool using several scenarios. 
\end{abstract}

%\begin{keyword}
%	Conformance testing \sep Generating test suites \sep Complete test suites \sep Asynchronous systems \sep IOLTS
%\end{keyword}

%\keywords{model-based testing \and conformance checking \and test generation \and reactive systems \and automatic tool.}
% !TeX spellcheck=en_US
\section{Introduction}

Several real-world systems are characterized by reactive behaviors that interact constantly  with the environment by receiving input stimuli and producing outputs in response. 
Systems of this nature, in general, are also critical thus requiring a precise and  an automatic support in the development process.
Model-based testing methods and their respective tools have been largely applied in the testing activity on their system development process. 
The Input Output Labeled Transition System (IOLTS) models~\cite{tretmans2008,tretmans1999} %simao2014,
has been commonly employed as the formalism on testing asynchronous reactive systems.
In this setting, IOLTS models specify desired behaviors of an implementation candidate and the testing task aims to find faults in Implementations Under Test (IUTs).  

One important issue of model-based testing is conformance checking where we can verify whether a given IUT complies with its respective specification according to a certain fault model. 
Here we treat the classical notion of Input Output Conformance Testing (\ioco)~\cite{tretmans2008} and  a more recent conformance based on regular languages~\cite{bonifacio2018} to define fault models. 
The test suite generation is also deemed important specially when generating test cases for reactive systems in black-box setting. 
In this work, we present an automatic tool that can check conformance between a specification and  an IUT using both conformance relations. 
Our tool also can generate test suites based on their specifications modeled by IOLTSs and provide a black-box testing run.

We claim that \everest has a wider range of applications when compared to other tools of the literature~\cite{jtorx,tgv,tgv1} since it implements both the classical \ioco relation and the language-based conformance using regular languages.   
Real world and practical scenarios are described to show aspects related to both approaches, where 
%of test generation and conformance checking to show when a 
faults can be found using the language-based conformance 
but the \ioco relation method cannot detect it. 
%these faults could not be detected using the classical \ioco relation. 
%We also run 
Moreover, some experiments are performed to evaluate \everest and compared to a well-known tool from the literature under the conformance checking task. 
Furthermore, we evaluate our tool when generating and running 
test suites in a black-box scenario. 

The remainder of this paper is organized as follows. 
Section~\ref{sub:metodo} describes the conformance checking methods using regular languages and the classical \ioco relation, besides the approach to generate test suites. 
In Section~\ref{sub:ferramenta} we present our tool, describe some case studies and discuss important aspects comparing to another tool from the literature. 
%Section~\ref{sec:comparative} gives a comparative analysis between our tool and a tool from the literature. 
Practical experiments of conformance checking and test suite generation are given in Section~\ref{sec:experiments} to evaluate the tool's performance. 
Section~\ref{sub:conclusao} offers some concluding remarks and future directions. 

% !TeX spellcheck=en_US
%\vspace*{-1ex}
\section{A Model-Based Testing Method}\label{sub:metodo}

%We deal with a more recent testing 
%method for asynchronous reactive systems using IOLTS models. 
%The next subsections describe two  important issues of model-based testing and relate them to the classical ioco conformance~\cite{tretmans2008}. 
Asynchronous reactive systems are commonly  specified by IOLTS models. 
An IOLTS is a variation of a Labeled Transition System (LTS)~\cite{tretmans1992} %,daca2014,zeng2009,cartaxo2007
with the partitioning of input and output labels. 
\begin{definition}%(\cite{tretmans2008, simao2014}) 
	\label{def:iolts}
	An IOLTS $\yS$ is given by $(S, s_0, L_I, L_U, T)$ where: 
	$S$ is the set of states;
	$s_0 \in S$ is the initial state;
	$L_I$ is a set of input labels;
	$L_U$ is a set of output labels;
	$L=L_I \cup L_U$ and $L_I \cap L_U =\emptyset$;
	$T$ $\subseteq S \times (L \cup \{ \tau \}) \times S$ is a finite set of transitions, %\cam{será que precisa deixar isto sobre ações internas?} 
	where the internal action $\tau \notin L$; and 
	$(S, s_0, L, T)$ is the underlying LTS associated with $\yS$.	
	%	$\yI = (Q,q_0,L_I,L_U, R)$, where:
	%	\begin{itemize}
	%		\item $S$ is the set of states;
	%		\item $s_0 \in S$ is the initial state;
	%		\item $L_I$ is a set of input labels;
	%		\item $L_U$ is a set of output labels;
	%		\item $L=L_I \cup L_U$ and $L_I \cap L_U =\emptyset$;
	%		\item $T$ $\subseteq S \times (L \cup \{ \tau \}) \times S$ is a finite set of transitions, where the internal action $\tau \notin L$; and 
	%		\item $(S, s_0, L, T)$ is the underlying LTS associated with $\yS$.	
	%	\end{itemize}	
\end{definition}

%%%%%%% veio do ICSEA 
We indicate by $(s, l, r)\in T$ a transition of an IOLTS/LTS model from state $s \in S$ to state $r \in S$ with the label $l \in (L \cup \{ \tau \})$. 
A transition $(s, \tau, r)\in T$ indicates an internal action, which means that an external observer cannot see the movement from state $s$ to state $r$ in the model. 

An IOLTS may also have quiescent states.  
A state $s$ is quiescent if no output $x\in L_U$ and an internal action $\tau$ are defined on it~\cite{tretmans2008}. 
When a state $s$ is quiescent a transition $(s,\delta,s)$ is added to $T$, where $\delta \notin L_ \tau$. 
Note that $L \cup \{\tau \}$ is denoted by $L_\tau$ to ease the notation.
We also note that in a real black-box testing scenario where an IUT sends messages to the tester and receives back responses, quiescence will indicate that the IUT can no longer respond to the tester, or it has timed out, or even it is simply slow~\cite{bonifacio2018}. 

We also introduce the notion of paths that will be useful to define semantics over IOLTS/LTS models. 
\begin{definition}(\cite{bonifacio2018})
	Let $\yS = (S, s_0, L, T)$ be a LTS and $p,q \in S$. 
	Let $\sigma = l_1, \cdots, l_n$ be a word in $L^\star_\tau$. 
	We say that $\sigma$ is a \emph{path} from $p$ to $q$ in $\yS$ if there are states $r_i \in S$, and labels $l_i \in L_\tau$, $1 \leq i \leq n$, such that $(r_{i-1}, l_i, r_i) \in T$, with $r_0=p$ and $r_n = q$.		 
	We  say that $\alpha$ is an \emph{observable path} from $p$ to $q$ 
	in $\yS$ if we remove  the internal actions $\tau$ from $\sigma$. 
\end{definition}
A path can also be denoted by $s \xrightarrow[]{\sigma} s'$, where the behavior $\sigma \in L^\star_\tau$  starts in the state $s \in S$ and reaches the state $s' \in S$. 
An observable path $\sigma$, from $s$ to $s'$, is denoted by $s\xRightarrow[]{\sigma} s'$. 
We can also write $s \xrightarrow[]{\sigma} $ or $s\xRightarrow[]{\sigma} $ when the reached state is not important. 
We call by \emph{paths of $s$} all those paths that start at the state $s$. 

Now we give  the notion of semantics over IOLTS/LTS models.
%The semantics of an LTS model is given by the paths that start from their initial state. 
\begin{definition}\label{def:traceSemantica}(\cite{bonifacio2018}).
	Let $\yS = (S, s_0, L, T)$ be a LTS  and $s \in S$:
	(1) The set of paths of $s$ is given by $tr(s) =  \{\sigma \arrowvert s \xrightarrow[]{\sigma} \} $ and the set of observable paths of $s$ is 
	$otr(s) = \{ \sigma \arrowvert s \xRightarrow[]{\sigma} \} $. 
	(2) The semantics of $\yS$ is $tr(s_0)$ or $tr(\yS)$ and the observable semantics of $\yS$ is $otr(s_0)$ or $otr(\yS)$.
\end{definition}
The semantics of an IOLTS is defined by the semantics of the underlying LTS.

\subsection{Conformance Checking}

Given an IOLTS~\cite{tretmans2008} %tretmans1999,simao2014,
specification, a conformance checking task can determine whether an IUT complies with the corresponding specification according to a specific fault model. 
The classical \ioco~\cite{tretmans2008,tretmans1999} relation establishes a notion of  conformance when input stimuli are applied to both the specification and the IUT, and observing if outputs produced by the IUT are also defined in the specification model~\cite{bonifacio2018,tretmans2008}.
\begin{definition}(\cite{tretmans2008}).
	Let $\yS = (S, s_0, L_I, L_U, T)$ be a specification and let $\yI = (Q, q_0, L_I, L_U, R)$ be an IUT,  %, ambas IOLTS. 
	$ \yI \; \ioco \; \yS$  if, and only if, $out(q_0 \; after \; \sigma) \subseteq out(s_0 \; after \; \sigma)$ for all $ \sigma \in otr(\yS)$, where $s \; after \; \sigma = \{q| s \xRightarrow[]{\sigma} q\}$ for every $s \in S$.
	Otherwise, 	$ \yI$ \sout{\ioco\!\!} $\yS$.
\end{definition}

A more recent conformance relation~\cite{bonifacio2018} has also been proposed using regular languages. 
Given an IUT $\yI$, a specification $\yS$, and regular languages $D$ and  $F$, $\yI$ complies with $\yS$ according $(D, F)$, i.e, $\yI  \; conf_{D,F} \;\yS$ if, and only if, no undesirable behavior of $F$ is observed in $\yI$ and is specified in $\yS$, and all desirable behaviors of $D$ are observed in $\yI$ and also are specified in $\yS$.
\begin{definition}(\cite{bonifacio2018})\label{definicao:confLing}
	Given an alphabet $L=\L_I \cup L_U$ and languages $\mathcal{D,F} \subseteq L^*$ over $L$. 
	Let $\yS$ and  $\yI$ be IOLTS models over $L$ we have that $\yI\; conf_{D,F}\; \yS $ if, and only if, 
	(i) $ \sigma \in otr(\yI) \cap F$,  then $\sigma \notin otr(\yS)$; 
	(ii) $ \sigma \in otr(\yI) \cap D$, so  $\sigma \in otr(\yS)$.
\end{definition}
Proposition~\ref{prop:verifConf} establishes this new notion with a wider fault coverage where desirable and undesirable behaviors can be specified by regular languages. 
\begin{prop}\label{prop:verifConf}(\cite{bonifacio2018}).
	Let $\yS$ and $\yI$ be IOLTS models over an alphabet $L=L_I\cup L_U$, and languages $D,F \subseteq L^*$ over $L$. 
	we say that $\yI \;conf_{D,F}\; \yS$ if, and only if, 
	$otr(\yI) \cap [(D \cap \overline{otr}(\yS)) \cap (F \cap otr(\yS))] = \emptyset$, where $\overline{otr}(\yS) = L^* - otr(\yS)$.
\end{prop}

Both notions of conformance can be related by the following lemma, where the language-based conformance given in Definition~\ref{definicao:confLing} restrains the classical \ioco relation.  %~\ref{lema:iocoLing}. 
\begin{lemma}\label{lema:iocoLing}(\cite{bonifacio2018}).
	Let $\yI = (Q, q_0, L_I, L_U, R)$ be an IUT and let $\yS = (S, s_0, L_I, L_U, T)$ be a specification, we say that $\yI \; \ioco \; \yS$ if, and only if,  $\yI \; conf_{D,F} \; \yS$ when $D=otr(\yS)L_U$ and  $F=\emptyset$. 
\end{lemma}

%%%%%%%%%%%%%%%%%%

Bonifacio and Moura~\cite{bonifacio2018} have proposed a conformance checking based on automata theory~\cite{sipser2006}. 
LTS/IOLTS models are transformed into Finite State Automatons~(FSAs).  %and apply union, intersection, and complement operations over regular languages. 
%An FSA is formally given by $\yA = (S, s_0, L, T, F)$, where $\yS = (S, s_0, L, T)$ is the underlying LTS associated with $\yA$. 
%Note that the set of final states in $\yA$ is defined by all states of $\yS$, i.e., $F = S$.
%Since the 
The semantics of an FSA is then given by the language it accepts, and $R\subseteq L^\star$ is \emph{regular} if there is an FSA $\mathcal{M}$  such that $L(\mathcal{M})=R$, where $L$ is an alphabet~\cite{sipser2006}.  
Therefore we can effectively construct the  the automatons  $\yA_{D}$  and $\yA_{F}$  such that $D$ and $F$ are regular languages and $D = L(\yA_D)$ and $F = L(\yA_F)$. 

The notions of the test case and test suite are also defined according to regular languages. % are given as follows. 
\begin{definition}(\cite{bonifacio2018}).
	Let $L$ be a set of symbols, a test suite $T$ over $L$ is a language, where $T \subseteq L^\star$, so that each $\sigma \in T$ is a test case.
\end{definition}
If the test suite is a regular language, then there is an  FSA $\mathcal{A}$ that accepts it, such that the final states of  $\mathcal{A}$ are fault states.
Further a set of undesirable behaviors can be defined by the fault states. 
We call this set by \emph{fault model} of $\yS$~\cite{bonifacio2018}.

Hence we can obtain a complete test suite for an IOLTS specification $\yS$ and a pair of languages $(D,F)$ using Proposition~\ref{prop:verifConf}. 
The test suite $T = [(D\cap \overline{otr}(\yS)) \cup (F \cap otr(\yS))]$ is able to detect the absence of desirable behaviors specified by $D$ and the presence of undesirable behaviors specified by $F$ in the specification $\yS$. 
We declare that an IUT $\yI$ complies with a specification $\yS$   
if there is no test case of the test suite $T$ that is also a behavior of $\yI$~\cite{bonifacio2018}. 

%We also provide the determinization of models which is useful in this method. 
%Therefore, from a deterministic 
In the process we first obtain an automaton $\yA_1$ induced by the IOLTS specification $\yS$. 
%IOLTS $\yS$ we can obtain the  that is also deterministic.  
Since $L(\yA_1) = otr(\yS)$ we can effectively construct an FSA $\yA_2$ such that $L(\yA_2) = L(\yA_F) \cap L(\yA_1) = F \cap otr(S)$. 
Also, consider the FSA $\yB_1$ obtained from $\yA_1$ by reversing its set of final states, that is, a state $s$ is a final state in $\yB_1$ if, and only if, $s$ is not a final state in $\yA_1$. 
Clearly, $L(\yB_1) = \overline{L(\yA_1)} = \overline{otr}(\yS)$. 
We can now get an FSA $\yB_2$ such that $L(\yB_2) = L(\yA_D) \cap L(\yB_1) = D \cap \overline{otr}(\yS)$.
Since $\yA_2$ and $\yB_2$ are FSAs,  we can construct  an FSA $\yC$  such that $L(\yC) = L(\yA_2) \cup L(\yB_2)$, where  $L(\yC) = T$.
We can conclude that when $D$ and $F$ are regular languages and $\yS$ is a deterministic specification, then a complete FSA $\mathcal{T}$ can be constructed such that $L(\mathcal{T}) = T$. 

Next proposition states an algorithm with a polynomial time complexity for the language-based verification. 
\begin{prop}(\cite{bonifacio2018})\label{prop:2}
	Let $\yS$ and $\yI$ be the deterministic specification and implementation  IOLTSs over $L$
	with $n_S$ and $n_I$ states, respectively. Let also $|L| = n_L$. Let $\yA_D$ and $\yA_F$ be deterministic FSAs
	over $L$ with $n_D$ and $n_F$ states, respectively, and such that $L(\yA_D) = D$ and $L(\yA_F) = F$. Then,
	we can effectively construct a complete FSA $\mathcal{T}$ with $(n_S + 1)^2n_Dn_F$ states, and such that $L(\mathcal{T})$
	is a complete test suite for $\yS$ and $(D, F)$. 
	Moreover, there is an algorithm, with polynomial time
	complexity $\Theta(n^2_Sn_In_Dn_Fn_L)$ that effectively checks whether $\yI conf_{D,F} \yS$ holds.
\end{prop}

Theorem~\ref{theo} shows that we can obtain a similar result for the \ioco relation using Lemma~\ref{lema:iocoLing}. 
\begin{theorem}(\cite{bonifacio2018})\label{theo}
	Let $\yS$ and $\yI$ be deterministic specification and implementation IOLTSs over $L$ with
	$n_S$ and $n_I$ states, respectively. Let $L = L_I \cup L_U$, and $|L| = n_L$. Then, we can effectively construct
	an algorithm with polynomial time complexity $\Theta(n_Sn_In_L)$ that checks whether $\yI$ \ioco $\yS$ holds.
\end{theorem}

%%%%%%%%%%%%%%%%%%%
\subsection{Test Suite Generation}

In this work we also provide the test suite generation in a black-box testing setting using the notion of Test Purposes (TPs)~\cite{tretmans2008,bonifacio2018}. 
A TP is formally defined by an IOLTS with two special states $\{pass, fail\}$ and, in practice, it represents an external tester that interacts  with an IUT. 
Thus a fault model is composed by a set of TPs that are derived from a given specification.
\begin{definition}(\cite{bonifacio2018})
	Let $L_I$  and $L_U$ be the input and output alphabets, respectively, with $L = L_I \cup L_U$.  
	A Test Purpose (TP) over $L$ is defined by an IOLTS $\yT \in \mathcal{IO}(L_U,L_I)$ such that for all $\sigma \in L^*$ does not hold $fail \xRightarrow[]{\sigma} pass$ and $pass \xRightarrow[]{\sigma} fail$. 
	The fault model over $L$ is the finite set of TPs over $L$.
\end{definition}
To ease the notation from now on we will denote  by $\mathcal{IO}(L_I, L_U)$ the class of all IOLTSs over $L=L_I \cup L_U$.

The test case generation proposed by Tretmans~\cite{tretmans2008}, based on \ioco relation, imposes some restrictions over the formal models. 
All TPs must be acyclic, with a finite run, and input-enabled,  since the tester cannot predict the output produced by a black-box IUT. 
Therefore, all output actions that are produced by the IUT must be enabled in the respective TP.
Moreover, they must be output-deterministic, \emph{i.e.} each state can send only one output symbol to the IUT in order to avoid arbitrary and non-deterministic choices. 
In the $pass$ and $fail$ states only self-loop transitions are allowed since verdicts are obtained in these states. 
\begin{definition}(\cite{bonifacio2018})
	Let $\yS \in \mathcal{IO}(L_I, L_U)$. 
	We say that $\yS$ is output-deterministic if $|out(s)| = 1$ and $\yS$ is input-enabled if $inp(s) = L_I$ for all $s \in S$, where $out(s)$ and $inp(s)$ give outputs and inputs, respectively, defined at state $s$.
	% A classe de todos os IOLTSs input-enabled sobre os alfabetos $L_I$ e $L_U$ é simbolizada por $\mathcal{IOE}(L_I, L_U)$.
\end{definition}

%Um modelo de falhas ioco completo cujos os TPs são \textit{input-enabled} pode então ser construído conforme a Proposição~\ref{propMF}.
%\begin{prop}(\cite{bonifacio2018})\label{propMF}
%	Para qualquer especificação $\yS$ é possível 
%	se construir 
%	um modelo de falhas $\mathcal{M}$ que seja ioco completo, determinístico, \textit{input-enabled}, com um único estado \textit{fail} e nenhum estado \textit{pass}.
%\end{prop}

Hence all restrictions imposed by Tretmans~\cite{tretmans2008} are satisfied when a TP is input-enabled, output-deterministic and acyclic except for $pass$ and $fail$ states. 
However we see that a bound over the number of states to be considered in the IUTs must be imposed to keep the TP acyclic in practice. 
So the test suite completeness property is guaranteed if given an IUT $\yI$ and a specification $\yS$, $\yI \; ioco \; \yS$ for all IUT that conforms to $\yS$. 
Otherwise we say that  $\yI$ \sout{ioco} $\yS$. 
Therefore we define a class of implementations to guarantee the 
\ioco completeness property for test suite generation establishing an upper bound on the number of states of the IUTs.
%\begin{definition}(\cite{bonifacio2018})
%	Let $\mathcal{IMP} \subseteq \mathcal{IO}(L_I,L_U)$ and $m \geq 1$, 
%	$\mathcal{IMP}[m]$ represents the subclass of IOLTS in $\mathcal{IMP}$ composed of all models with at most $m$ states. 
%	Let $\yS \in \mathcal{IO}(L_I,L_U)$, the fault model $\mathcal{M}$, over $L_I \cup L_U$, is $m$-\ioco-complete for $\yS$ relatively to  $\mathcal{IMP}$ if, and only if,  $\mathcal{M}$ is \ioco-complete to $\yS$ relatively to $\mathcal{IMP}[m]$. 
%	In this case, all TPs are deterministic and acyclic.
%\end{definition}

%Whereas the construction of a \ioco-complete fault model is not possible, in this work $m$-\ioco-complete fault models are built, where $m$ is the maximum number of IUT states.

%um parâmetro fornecido. 
%\begin{prop}(\cite{bonifacio2018})\label{prop:multiGrafo}
%	Dado o IOLTS determinístico $\yS = (S,s_0,L_I, L_U,T) \in \mathcal{IO}(L_I,L_U)$ e $m \geq 1$. Então existe um modelo de falhas $\mathcal{M}$ que é $m$-ioco completo para $\yS$, o qual todos os TPs de $\mathcal{M}$ são acíclicos e determinísticos.
%\end{prop}

Now we are in position to construct a complete test suite using the notion of TPs. 
But first we generate a multigraph structure as proposed by Bonifacio and Moura~\cite{bonifacio2018}. 
So given a IUT $\yI$ and a specification $\yS$, we remark that $m$ is the bound over the number of states to be considered on the IUT, and $n$ is the number of states in $\yS$. 
Then the multigraph must have $mn + 1$ levels, and at each level if a transition of $\yS$ gives rise to a cycle then we must create a transition to states on next level of the multigraph. 
The $fail$ state is also added as well as transitions to $fail$ which are not defined for all $l \in L_U$, and every state of the multigraph. 

Having an acyclic multigraph at hand then we can extract TPs using a simple breadth-first search algorithm from the initial state to $fail$. 
We can guarantee the input-enabledness property by adding the \textit{pass} state to the TP and, for every output of $L_U$  and all states, we add transitions to the $pass$ state where the output is not defined. 
Self-loops labeled by each $l \in L_U$ are also added to the $pass$ and $fail$ states. 
The output-deterministic property  is also obtained by adding a transition 
from each state where an input is not defined with any input of $L_I$ to the $pass$ state.  
Note that we always refer to an input symbol of $L_U$ or an output symbol of $L_I$ from the perspective of the IUT, as commonly denoted in the literature~\cite{tretmans2008,bonifacio2018}. 

%\begin{prova}
%Considere qualquer propósito de teste $\yTP=(S_{\yT},s_{0,0},L_U,L_I, T_{\yT}) \in \mathcal{M}$ e qualquer $t \in S_{\yT}$. Para assegurar a propriedade \textit{input-enabledness}, o estado \textit{pass} é adicionado em $S_{\yT}$. Para cada $l \in L_U$, se não existe uma transição $(t, l, t') \in T_{\yT}$ para qualquer $t' \in S_{\yT}$ e $t \neq fail$ é adicionado a transição $(t, l, pass) \in T_{\yT}$. Para todo $l \in L_U$ é adicionado também as transições $(pass, l, pass)$ e $(fail,l,fail)$ em $T_{\yT}$. Depois destes ajustes o TP é determinístico,  \textit{input-enabled} e acíclico exceto pelos \textit{self-loops} nos estados \textit{pass} e \textit{fail}.
%
%Para garantir que o TP seja \textit{output-deterministic}, é necessário ter no máximo uma transição $(t, l, t')$ em $\yTP$ para cada $l \in L_I \cup L_U$, como na etapa anterior não foi adicionado nenhuma transição rotulada por $L_I$, se $l \in L_I$ então $t$ já é  \textit{output-deterministic}, se $l \notin L_I$ então é escolhido qualquer símbolo $l' \in L_I$ e a transição $(t, l', pass)$ é adicionada em $T_{\yT}$. 
%\end{prova}

Next the test run is defined based on the synchronous product between a TP $\yT$ and an IUT $\yI$, denoted by $\yI \times \yT$.
The TP interacts with the IUT producing outputs that are sent to $\yI$ as inputs. 
Likewise, the IUT receives the actions and produces outputs that are sent to $\yT$ as inputs. 
Therefore, the output alphabet  of $\yT$ corresponds to $L_I$, the input alphabet of the IUT, and the input  alphabet of $\yT$ corresponds to $L_U$, the output alphabet of the IUT, as mentioned above. 
\begin{definition}(\cite{bonifacio2018})
	Let an IUT $\yI =(S_{\yI}, q_0, L_I,L_U, T_{\yI}) \in \mathcal{IO}(L_I,L_U)$ and the TP $\yT =(S_{\yT}, q_0, L_U,L_I, T_{\yT})) \in \mathcal{IO}(L_U,L_I)$. 
	We say that $\yI \; \text{passes} \; \yT$ if for any $\sigma \in (L_I, L_U)^*$ and any state $q \in S_{\yI}$, we do not have $(t_0,q_0) \xRightarrow[]{\sigma} (fail, q)$ in $\yI \times \yT$. 
	A path can be denoted by $q_0 \xRightarrow[]{\sigma} q$  where the behavior $\sigma$ starts in the state $q_0$ and reaches the state $q$.	
	Let $\mathcal{M}$ be the fault model, we say that $\yI \; \text{pass} \; \mathcal{M}$, if $\yI$ passes all TPs in $\mathcal{M}$. 
	Then given an IOLTS $\yS$ and a set 
%	of IOLTS 
	$\mathcal{IMP} \subseteq \mathcal{IO}(L_U,L_I)[m]$,  
	we say that $\mathcal{M}$ is  $m$-\ioco\!\!-complete to $\yS$ concerning $\mathcal{IMP}$ if for all implementation $\yI \in \mathcal{IMP}$ we have $\yI \; ioco \; \yS$ if, and only if, $\yI \; \text{passes} \; \mathcal{M}$. 
\end{definition}

The verdicts are obtained when TPs reach the special states. 
The \textit{fail} verdict gives rise to 
a fault behavior whereas the \textit{pass} verdict 
denotes a desirable behavior. 
Further details can be found in~\cite{bonifacio2018,icsea2019}. 
%A fault behavior is obtained when \textit{fail} state is reached, 
%and acceptable behavior when the \textit{pass} state is reached~\cite{bonifacio2018}. 

%If an IUT $\yI = (Q,q_0,L_I, L_U, R)$ does not pass by the fault model $\mathcal{M}$, we have that $(t_0,q_0) \xRightarrow[]{\sigma} (fail, q)$ in $\mathcal{M} \times \yI$, for some $\yTP = (S_{\yT},t_0,L_U,L_I, T_{\yT})$ in $\mathcal{M}$. 
%So $t_0 \xRightarrow[]{\sigma} fail$ in $\yTP$ and $q_0 \xRightarrow[]{\sigma} q$ in $\yI$.

%\begin{definition}(\cite{bonifacio2018})
%	Given the set of input symbols $L_I$ and output $L_U$, with $L = L_I \cup L_U$, where $\mathcal{IO}(L_I,L_U)$ is the class of all IOLTS over alphabets $L_I$ and $L_U$.
%	A TP over $L$ is any IOLTS $\yT \in \mathcal{IO}(L_U,L_I)$ such that for all $\sigma \in L^*$ does not occur $fail \xRightarrow[]{\sigma} pass$ and $pass \xRightarrow[]{\sigma} fail$. 
%	The expression $\xRightarrow[]{\sigma}$ denotes the path to process the word $\sigma$.
%	A fault model over $L$ is a finite set of TPs over $L$.
%\end{definition}

Finally, next proposition determines a fault model that is composed by TPs obtained from a multigraph which, in turn, is constructed based on the corresponding specification. 
\begin{prop}\label{prop:tp}
	Let the deterministic IOLTS $\yS \in \mathcal{IO}(L_I,L_U)$ and $m \geq 1$. So there is a fault model $\mathcal{M}$ that is $m$-\ioco\!\!-complete for $\yS$ relatively to  $\mathcal{IO}(L_I,L_U)[m]$, IOLTSs at most $m$ states, whose TPs are deterministic, output-deterministic, input-enabled and acyclic except for self-loops on 
	%\cam{add 'the'?} 
	\textit{pass} and \textit{fail} states.
\end{prop}

% !TeX spellcheck=en_US
\section{A Testing Tool for Reactive Systems}\label{sub:ferramenta} 

We have developed the \everest\!\footnote{\textit{conformancE Verification on tEsting ReactivE SysTems}}\!~\cite{icsea2019} tool to check conformance, generate test suites and run those testes over reactive systems specified by LTS/IOLTS models. 
% and according to the foundation given in Section~\ref{sub:metodo}.  
%The detailed process on checking conformance can be found in~\cite{icsea2019,DBLP:journals/corr/abs-1905-08914}.
Our tool is settled down on four main modules: 
configuration; \ioco conformance; language-based conformance; and test generation/run. 
When checking conformance if an IUT does not conform to the specification our tool is able to yield the verdict along with the paths induced by the test cases which have detected the corresponding faults. 
%In language-based conformance verification view, \textit{Desirable} and \textit{Undesirable} behaviors  must be specified by regular expressions. 
%The Kleene closure~\cite{sipser2006} is assumed over the alphabet, when no regular expression is provided, to identify faults on models are not isomorphic.
In the test generation/run module we can generate multigraphs as well as TPs according to the respective specification, and also allow us to run a test suite over a given IUT. 

Next we describe a general case study to compare the conformance checking process using \everest and JTorx, a well-known testing tool from the literature~\cite{belinfante2014}. 
Further we demonstrate in practice the test suite generation process using our tool and stand out some aspects related to the language-based testing approach. 
In the sequel we give a real-world case study of an Automatic Teller Machine (ATM) to explore some real scenarios using both testing tools. 
Finally we describe a comparative analysis by means different aspects between \everest and JTorx. 

% !TeX spellcheck=en_US
\subsection{Conformance checking  process}

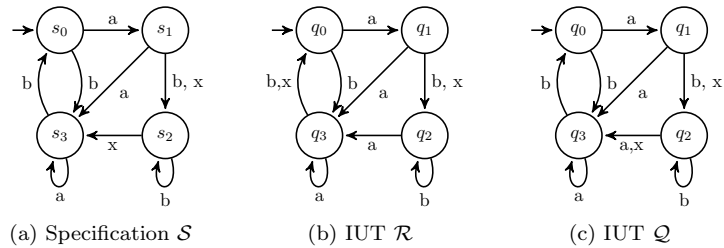
\begin{figure}[hbt]
%\vspace*{-6ex}
	\begin{center}
		\begin{tabular}{@{}ccc@{}}	
				\subfloat[Specification $\yS$
			%Model underlying $.aut$. 
			\label{fig:SubjAut}]{
				\begin{tikzpicture}[->,>=stealth',shorten >=1pt,auto,node distance=2cm, semithick,initial text =,scale=0.7,transform shape]
				
				\node[initial, state]       (s0) {$s_0$};
				\node[state]                (s3) [below of=s0]{$s_3$};
				\node[state]                (s1) [right of=s0] {$s_1$};
				\node[state]               	(s2) [below of=s1]{$s_2$};
				
				\path 
				(s0) edge               node {a}    	(s1)
				edge  [bend left]	node {b}   	(s3)
				(s1) edge               node {a}   	(s3)
				edge               node {b, x}   (s2)
				(s2) edge               node {x}   	(s3)
				edge  [loop below] node {b}   	(s2)   
				(s3) edge  [bend left]  node {b}   	(s0)
				edge  [loop below] node {a}   	(s3);
				\end{tikzpicture}
				
			}\quad &	
			\subfloat[][IUT $\yR$ \label{fig:imp-quase-iso-modificada}]{
				\begin{tikzpicture}[->,>=stealth',shorten >=1pt,auto,node distance=2cm, semithick,initial text =,scale=0.7,transform shape]
				
				\node[initial, state]       (s0) {$q_0$};
				\node[state]                (s3) [below of=s0]{$q_3$};
				\node[state]                (s1) [right of=s0] {$q_1$};
				\node[state]               	(s2) [below of=s1]{$q_2$};

				\path 
				(s0) edge               node {a}    	(s1)
				edge  [bend left]	node {b}   	(s3)
				(s1) edge               node {a}   	(s3)
				edge               node {b, x}   (s2)
				(s2) edge               node {a}   	(s3)
				edge  [loop below] node {b}   	(s2)   
				(s3) edge  [bend left]  node {b,x}   	(s0)
				edge  [loop below] node {a}   	(s3);
				\end{tikzpicture}
			}\quad
			&
			\subfloat[][IUT $\yQ$ \label{fig:implementacao_EC_IOCO}]{
				\begin{tikzpicture}[->,>=stealth',shorten >=1pt,auto,node distance=2cm, semithick,initial text =,scale=0.7,transform shape]
				
				\node[initial, state]       (q0) {$q_0$};
				\node[state]                (q3) [below of=q0]{$q_3$};
				\node[state]                (q1) [right of=q0] {$q_1$};
				\node[state]               	(q2) [below of=q1]{$q_2$};

				\path 
				(q0) edge               node {a}    	(q1)
				edge  [bend left]	node {b}   	(q3)
				(q1) edge               node {a}   	(q3)
				edge               node {b, x}   (q2)
				(q2) edge               node {a,x}   	(q3)
				edge  [loop below] node {b}   	(q2) 
				(q3) edge  [bend left]  node {b}   	(q0)
				edge  [loop below] node {a}   	(q3);
				\end{tikzpicture}
			}	
			
		\end{tabular}
	\end{center}
%\vspace*{-5ex}
	\caption{IOLTS Models \label{fig:EC} }
%\vspace*{-2ex}
\end{figure}

A conformance testing process is provided by \everest and also by JTorx tool. 
Here we have applied a general case study to explore characteristics from both tools. 
So let $\yS$ of Figure~\ref{fig:SubjAut} be a specification and let $\yR$ and $\yQ$ depicted in Figures~\ref{fig:imp-quase-iso-modificada} and~\ref{fig:implementacao_EC_IOCO}, respectively, be IUTs. 
Also assume $L_I=\{a,b\}$  and $L_U=\{x\}$. 

We first check if the IUT $\yR$ conforms to the specification 
$\yS$. 
Our tool yielded a verdict of non-conformance and has generated $T_1=\{b,aa,ba,aaa,ab,ax, abb, axb\}$ as a test suite. 
All test cases were induced by different paths that reach a fault and 
they were extracted using a transition cover strategy over the specification. 
The same verdict was obtained using JTorx for this same scenario, as expected, but it returned the test suite $T_2=\{b,ax,ab\}$. 
It is clear that $T_2\subseteq T_1$, \emph{i.e.}, JTorx has produced only one test case per fault in contrast to \everest thas has used transition cover. 
Hence notice that such a wider range of coverage provided by \everest can be more useful in a fault mitigation process. 

In a second scenario, we assume $\yQ$ as an IUT for the specification $\yS$. 
At this time no fault was detected by both tools under the classical \ioco relation. 
However, \everest was able to find a fault under the language-based conformance relation, where desirable behaviors were specified by the regular language $D=(a|b)^*ax$ and  no undesirable behavior was defined, so $F=\emptyset$. 
The set $D$ denotes behaviors that are induced by paths finishing with an input action $a$ followed by an output $x$ produced in response. 
A verdict of non-conformance was then obtained by our tool revealing a fault detected by the test suite $T=\{ababax,abaabax\}$. 
We remark that JTorx which implements only the classical \ioco relation was not able to detect this fault. 
Again, it is clear that \everest is more general and can be applied to a wider range of scenarios when compared to the JTorx. 

\subsection{\everest test suite generation}

\begin{figure}[tb]
%\vspace*{-5ex}
\center
\hspace*{-5ex}
\begin{tikzpicture}[font=\sffamily,node distance=1cm, auto,
%xscale=.6,yscale=.5,transform shape] %
scale=0.65,transform shape] %,rotate=90]

%linha 0
  \node[ initial by arrow, initial text={}, punkt] (s00) {$s_{0,0}$};
  \node[punkt, inner sep=3pt,right=2cm  of s00] (s10) {$s_{1,0}$};
  \node[punkt, inner sep=3pt,right=2cm  of s10] (s20) {$s_{2,0}$};
  \node[punkt, inner sep=3pt,right=2cm  of s20] (s30) {$s_{3,0}$};

%linha 1
  \node[punkt, inner sep=3pt,below=1.5cm of s00] (s01) {$s_{0,1}$};
    \node[punkt, inner sep=3pt,right=2cm  of s01] (s11) {$s_{1,1}$};
  \node[punkt, inner sep=3pt,right=2cm  of s11] (s21) {$s_{2,1}$};
  \node[punkt, inner sep=3pt,right=2cm  of s21] (s31) {$s_{3,1}$};

%linha 2
  \node[punkt,inv,inner sep=3pt,below=1.5cm of s01] (s02) {};
    \node[punkt, inv,inner sep=3pt,right=2cm  of s02] (s12) {};
  \node[punkt, inv,inner sep=3pt,right=2cm  of s12] (s22) {};
  \node[punkt, inv,inner sep=3pt,right=2cm  of s22] (s32) {};

%linha 15
  \node[punkt,inner sep=3pt,below=2.5cm of s02] (s015) {$s_{0,15}$};
    \node[punkt, inner sep=3pt,right=2cm  of s015] (s115) {$s_{1,15}$};
  \node[punkt, inner sep=3pt,right=2cm  of s115] (s215) {$s_{2,15}$};
  \node[punkt, inner sep=3pt,right=2cm  of s215] (s315) {$s_{3,15}$};

%linha 16
  \node[punkt,inner sep=3pt,below=1.5cm of s015] (s016) {$s_{0,16}$};
    \node[punkt, inner sep=3pt,right=2cm  of s016] (s116) {$s_{1,16}$};
  \node[punkt, inner sep=3pt,right=2cm  of s116] (s216) {$s_{2,16}$};
  \node[punkt, inner sep=3pt,right=2cm  of s216] (s316) {$s_{3,16}$};

% Fails

\node[punkts, invisible,inner sep=0pt,below right=0.7cm and .7cm of s00] (fail0) {$\yfail$};
\node[punkts, invisible,inner sep=0pt,below left=0.7cm and 0.1 of s20] (fail2) {$\yfail$};
\node[punkts, invisible,inner sep=0pt,below right=0.5cm and 1cm of s30] (fail3) {$\yfail$};

\node[punkts, invisible,inner sep=0pt,below right=0.7cm and .7cm of s015] (fail150) {$\yfail$};
\node[punkts, invisible,inner sep=0pt,below left=0.7cm and 0.1 of s215] (fail152) {$\yfail$};
\node[punkts, invisible,inner sep=0pt,below right=0.5cm and 1cm of s315] (fail153) {$\yfail$};

% transicoes do nivel 0 para nivel 1
    
\path (s00)    edge [pil]   	node[anchor=north,above]{a} (s10);
\path (s00)    edge [pil,bend left=35]   	node[anchor=south]{b} (s30);
\path (s00)    edge [pil]   	node[anchor=west]{ $\delta$} (s01);
 \path (s10)    edge [pil]   	node[anchor=north, above]{b,x} (s20);
%\path (s10)    edge [pil]   	node[anchor=west]{a} (s11);
 \path (s10)    edge [pil,bend left=25]   	node[anchor=south]{a} (s30);
\path (s20)    edge [pil]   	node[anchor=north, above]{x} (s30);
\path (s20)    edge [pil]   	node[anchor=west]{b} (s21);
%\path (s20)    edge [pil]   	node[anchor=north, above]{a} (s11);
\path (s30)    edge [pil]   	node[anchor=west]{a, $\delta$} (s31);
\path (s30)    edge [pil,out = 135, in = 480, out looseness=2, in looseness=1.6]   	node[anchor=north]{b} (s01); %%essa

% transicoes do nivel 1 para nivel 2
 
\path (s01)    edge [pil]   	node[anchor=north,above]{a} (s11);
\path (s01)    edge [pil,bend right=40]   	node[anchor=north]{b} (s31);
 \path (s11)    edge [pil]   	node[anchor=north, above]{b,x} (s21);
%\path (s11)    edge [pil,dashed]   	node[anchor=west]{a} (s12);
 \path (s11)    edge [pil,bend right=25]   	node[anchor=north]{a} (s31);
\path (s21)    edge [pil]   	node[anchor=north, above]{x} (s31);
%\path (s21)    edge [pil,dashed]   	node[anchor=south]{a} (s12);
\path (s31)    edge [pil,dashed]   	node[anchor=west]{a, $\delta$} (s32);
\path (s21)    edge [pil,dashed]   	node[anchor=west]{b} (s22);
\path (s01)    edge [pil,dashed]   	node[anchor=west]{b} (s02);
\path (s31)    edge [pil,dashed,bend left=30]   	node[anchor=north]{b} (s02); %%essa

% transicoes do nivel 15 para nivel 16
 
\path (s015)    edge [pil]   	node[anchor=north,above]{a} (s115);
\path (s015)    edge [pil,bend left=35]   	node[anchor=north]{b} (s315);
\path (s015)    edge [pil]   	node[anchor=west]{ $\delta$} (s016);
 \path (s115)    edge [pil]   	node[anchor=north, above]{b,x} (s215);
%\path (s115)    edge [pil]   	node[anchor=west]{a} (s116);
 \path (s115)    edge [pil,bend left=25]   	node[anchor=south]{a} (s315);
\path (s215)    edge [pil]   	node[anchor=north, above]{x} (s315);
\path (s215)    edge [pil]   	node[anchor=west]{b} (s216);
%\path (s215)    edge [pil]   	node[anchor=south]{a} (s116);
\path (s315)    edge [pil]   	node[anchor=west]{a, $\delta$} (s316);
\path (s315)     edge [pil,out = 135, in = 480, out looseness=1.5, in looseness=1.6]  	node[anchor=north]{b} (s016); %%essa
%[pil,out = 30, in = -50, in looseness=3, out looseness=2] 

% transicoes no nivel 16
 
\path (s016)    edge [pil]   	node[anchor=north,above]{a} (s116);
\path (s016)    edge [pil,bend right=35]   	node[anchor=north]{b} (s316);
 \path (s116)    edge [pil]   	node[anchor=north, above]{b,x} (s216);
 \path (s116)    edge [pil,bend right=25]   	node[anchor=north]{a, $\delta$} (s316);
\path (s216)    edge [pil]   	node[anchor=north, above]{x} (s316);

%transicoes fail

\path (s00)    edge [pil,dashed]   	node[anchor=north,above]{x} (fail0);
\path (s01)    edge [pil,dashed]   	node[anchor=west,below]{x} (fail0);
\path (s10)    edge [pil,dashed]   	node[anchor=north,above]{$\delta$} (fail0);
\path (s11)    edge [pil,dashed]   	node[anchor=north,above]{$\delta$} (fail0);
\path (s20)    edge [pil,dashed]   	node[anchor=west]{$\delta$} (fail2);
\path (s21)    edge [pil,dashed]   	node[anchor=east]{$\delta$} (fail2);
\path (s30)    edge [pil,dashed]   	node[anchor=west]{x} (fail3);
\path (s31)    edge [pil,dashed]   	node[anchor=west]{x} (fail3);

\path (s015)    edge [pil,dashed]   	node[anchor=north,above]{x} (fail150);
\path (s016)    edge [pil,dashed]   	node[anchor=west,below]{x, $\delta$} (fail150);
\path (s116)    edge [pil,dashed]   	node[anchor=west,below]{$\delta$} (fail150);
\path (s115)    edge [pil,dashed]   	node[anchor=west,below]{$\delta$} (fail150);
\path (s215)    edge [pil,dashed]   	node[anchor=west]{$\delta$} (fail152);
\path (s216)    edge [pil,dashed]   	node[anchor=east]{$\delta$} (fail152);
\path (s315)    edge [pil,dashed]   	node[anchor=west]{x} (fail153);
\path (s316)    edge [pil,dashed]   	node[anchor=west]{x,$\delta$} (fail153);

% dots
\path (s01)    edge [pil,invisible]   	node[]{\Large\bf $\vdots$} (s015);
\path (s31)    edge [pil,invisible]   	node[]{\Large\bf $\vdots$} (s315);
\end{tikzpicture}
%	\vspace*{-2ex}
\caption{A direct acyclic multi-graph D for specification $\yS$
}\label{fig:tp-spec-nocycle}
%\vspace*{-1ex}
\end{figure}
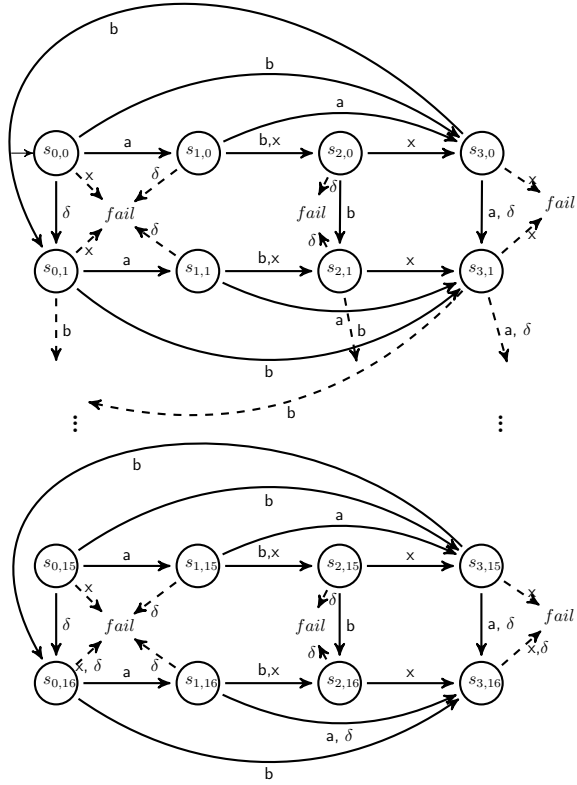

Now we turn into the test suite generation process provided by \everest tool. 
Assume again the specification $\yS$ of Figure~\ref{fig:SubjAut}. 
In the first step we construct direct acyclic multigraphs according to a given specification, as described in Section~\ref{sub:metodo}. 
The multigraph, partially depicted in Figure~\ref{fig:tp-spec-nocycle},  has four states at each level since the specification $\yS$ has four states ($n=4$). 
% !TeX spellcheck=en_US
All transitions connect nodes at the same level from left to right, or to the next level, so assuring the acyclic property. 
In this case we have considered IUTs with at most four states, \emph{i.e.} the same number of states that are in the specification ($m=n=4$). 
Therefore the number of levels in the multigraph is $mn+1=17$. 
Figure~\ref{fig:tp-spec-nocycle} shows the first two levels and also the two last levels of the multigraph. 
Note that we replicate the fail state in order not to clutter the figure.

\begin{figure}[hbt]
	%\vspace*{-2ex}
	\begin{center}
		\begin{tabular}{@{}cc@{}}			
			\subfloat[][TP induced by $aabbx$ \label{fig:tp-1}]{
				\begin{tikzpicture}[->,>=stealth',shorten >=1pt,auto,node distance=2cm, semithick,initial text =,scale=0.7,transform shape]
					
					\node[initial, state]       (s00) {$s_0,0$};
					\node[state]                (s10) [right of=s00]{$s_1,0$};
					\node[state]                (s30) [right of=s10] {$s_3,0$};
					\node[state]               	(s01) [right of=s30]{$s_0,1$};
					\node[state]               	(s31) [below of=s01]{$s_3,1$};
					\node[state]               	(fail) [below of=s31]{$fail$};
					\node[state]               	(pass) [left of=s31]{$pass$};

					\path 
					(s00) edge     node {a}    	(s10)
					(s00) edge     node {$b,\delta$}    	(pass)
					(s10) edge     node {a}    	(s30)
					(s10) edge     node {$b,\delta$}    	(pass)
					(s30) edge     node {b}    	(s01)
					(s30) edge     node {$a,\delta$}    	(pass)
					(s01) edge     node {b}    	(s31)
					(s01) edge     node {$a,\delta$}    	(pass)
					(s31) edge     node {x}    	(fail)
					(s31) edge     node {$a,b$}    	(pass)	
					(pass) 	edge  [loop below] node {$\delta, x$}   	(pass)
					(fail) 	edge  [loop right] node {$\delta, x$}   	(fail);
				\end{tikzpicture}
			}\\
%		\quad &
			\subfloat[][TP induced by $aaax$ \label{fig:tp-2}]{
				\begin{tikzpicture}[->,>=stealth',shorten >=1pt,auto,node distance=2cm, semithick,initial text =,scale=0.7,transform shape]
					
					\node[initial, state]       (s00) {$s_0,0$};
					\node[state]                (s10) [right of=s00]{$s_1,0$};
					\node[state]                (s30) [right of=s10] {$s_3,0$};
					\node[state]               	(s31) [below of=s30]{$s_3,1$};
					\node[state]               	(fail) [below of=s31]{$fail$};
					\node[state]               	(pass) [left of=s31]{$pass$};

					\path 
					(s00) edge     node {a}    	(s10)
					(s00) edge     node {$b,\delta$}    	(pass)
					(s10) edge     node {a}    	(s30)
					(s10) edge     node {$b,\delta$}    	(pass)
					(s30) edge     node {a}    	(s31)
					(s30) edge     node {$b,\delta$}    	(pass)	
					(s31) edge     node {x}    	(fail)
					(s31) edge     node {$a,b$}    	(pass)	
					(pass) 	edge  [loop below] node {$\delta, x$}   	(pass)
					(fail) 	edge  [loop right] node {$\delta, x$}   	(fail);
				\end{tikzpicture}
			}	
			
		\end{tabular}
	\end{center}
	%	\vspace*{-3ex}
	\caption{TPs from multigraph of Figure~\ref{fig:tp-spec-nocycle} \label{fig:TPs} }
	%	\vspace*{-5ex}
\end{figure}
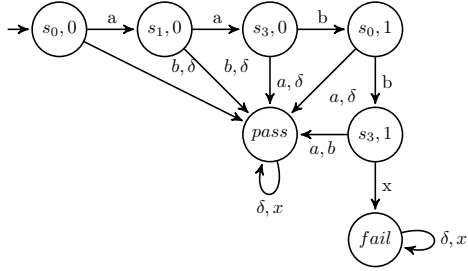
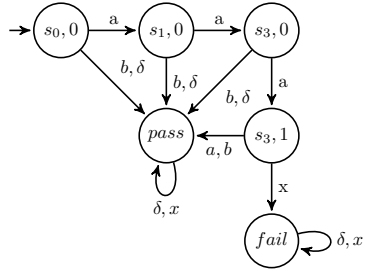

With a multigraph at hand we apply a breadth-first search algorithm to extract paths from the initial node $s_{0,0}$ up to the fail state. 
For instance, take $\alpha_1 = aabbx$. 
We see that $\alpha_1$ induces the path $s_{0,0} \rightarrow s_{1,0} \rightarrow s_{3,0} \rightarrow s_{0,1} \rightarrow s_{3,1} \rightarrow fail$. 
According to the Proposition~\ref{prop:tp} we can obtain a deterministic, acyclic, input-enabled and output-deterministic test purpose. 
So we obtain the test purpose depicted in Figure~\ref{fig:tp-1} based on $\alpha_1$ such that all these properties are satisfied. 
% !TeX spellcheck=en_US

The input-enabledness property is secured by adding a pass state and transitions from those states where no output is defined to the pass state. 
The construction is complete by adding self-loops to the pass and fail states labeled by all output actions. 
Regarding the output-determinism property, for every state that there is no input action defined on it we create a new transition form this state to the pass state with any input action.

For the sake of exemplification we also take the sequence $\alpha_2 = aaax$. 
In the same way we obtain the induced path over the multigraph and construct a corresponding  deterministic, acyclic, input-enabled and output-deterministic test purpose as depicted in Figure~\ref{fig:tp-2}. 
\everest has automatically constructed other 15 TPs based on paths induced by the set $\{\alpha_1,\alpha_2, x,a\delta,bx,\delta x,aax,bbx, ax\delta, ab\delta, \delta bx, b \delta x, aabx,bbbx,aa\delta x\}$ of sequences. 
% to detect a fault in the IUT $\yR$. 
For instance, the test suite $T=\{\alpha_1,\alpha_2,b, \delta, b\delta, bx, ab,a\delta, aa\delta, aaa, aab, aab\delta, aaba, aaaa, aaab, aaax\delta, \\aaaxx, aabba, aabbb, aabbx\delta, aabbxx\}$ has been generated using those two TPs depicted in Figure~\ref{fig:TPs}. 

After applying the test suite $T$ to the IUT $\yR$ a fault was detected. 
%, more precisely, using the test cases $\{\alpha_1, \alpha_2,bx, aaax\delta\}$.  
By a simple inspection we see that all test cases that lead $\yR$ from state $q_0$ to the same state $q_0$ can detect a fault when an output $x$ is produced at state $q_3$, and since $x$ is not defined at state $s_3$ of $\yS$. 
Our tool then returns a verdict of non-conformance reaching this fault which means that $\yR$ does not pass the test cases. 
So \everest declares that $\yR \; \ioco \; \yS $ does not hold.

% !TeX spellcheck=en_US
\subsection{A real-world case study}

Now we illustrate the conformance checking process by means a practical application using a real-world scenario. 
We specify an Automatic Teller Machine (ATM)~\cite{utting2007,naik2008} by an IOLTS with the input stimuli $L_I = \{ic,pin,acc,tra,sta,wd,amo\}$, and the output responses $L_U = \{cpi,bpi,mon,rec,ins, sho\}$. 

The intended meaning of the input actions are: $ic$, denotes the action when the user inserts his/her card into the ATM; 
$pin$, indicates the pin code was provided by the user; 
$tra$, requires the transfer amount; 
$acc$, indicates that a target account was provided; 
$sta$, requires an account statement; 
$wd$, denotes the user has requested a withdrawal; and 
$amo$, denotes the balance account. 
Similarly we have the meaning of output actions: 
$cpi$, says the pin code is correct; 
$bpi$, says the provided pin is wrong; 
$mon$, indicates the money was released; 
$rec$, indicates the receipt was provided to the user; 
$ins$, denotes an insufficient balance on account; and 
$sho$, indicates the statement was shown to the user.

\begin{figure}[hbt]
	\center
	\begin{tikzpicture}[->,>=stealth',shorten >=1pt,auto,node distance=2.5cm, semithick,initial text =,scale=0.7,transform shape]
		
		\node[initial, state]       (s0) {$s_0$};
		\node[state]                (s1) [right of=s0]{$s_1$};
		\node[state]                (s2) [right of=s1] {$s_2$};
		\node[state]               	(s3) [below of=s2]{$s_3$};	
		\node[state]               	(s7) [below of=s3]{$s_4$};
		\node[state]               	(s8) [below of=s1]{$s_5$};

		\path 
		(s0) edge               node {?ic}    (s1)	
		(s1) edge               node {?pin}   	(s2)
		(s2) edge               node {!cpi}   	(s3)
		edge  [bend right = 60]  node [above]{!bpi}   	(s1)
		(s3) edge   node {?wd}   	(s7)		 	
		(s7) edge   node {?amo}    (s8)	
		(s8) edge  [bend left = 30] node [below] {!mon} (s0)
		edge  [bend left = 30] node {!ins}   (s3);
	\end{tikzpicture}	
	\caption{ATM specification $\mathcal{A}$}
	\label{fig:spec-atm1}
\end{figure}
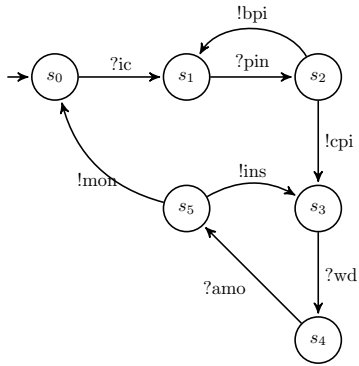

\begin{figure}[hbt]
	\center
	\begin{tikzpicture}[->,>=stealth',shorten >=1pt,auto,node distance=2.5cm, semithick,initial text =,scale=0.7,transform shape]
		
		\node[initial, state]       (s0) {$s_0$};
		\node[state]                (s1) [above of=s0]{$s_1$};
		\node[state]                (s2) [right of=s1] {$s_2$};
		\node[state]               	(s3) [right of=s2]{$s_3$};
		\node[state]       			(s4) [below of=s3]{$s_4$};
		\node[state]                (s5) [below of=s4]{$s_5$};
		\node[state]                (s6) [below of=s5] {$s_6$};
		\node[state]               	(s7) [below of=s2]{$s_7$};
		\node[state]               	(s8) [below of=s7]{$s_8$};
		\node[state]               	(s9) [left of=s1]{$s_9$};
		
		\path 
		(s0) edge               node {?ic}    (s1)		
		(s1) edge               node [below]{?pin}   	(s2)
		(s2) edge               node {!cpi}   	(s3)
		edge  [bend right = 60] node [above]{!bpi}   	(s1)   
		(s3) edge   node {?tra}   	(s4)
		edge [bend right = 50]  node [above]  {?sta}  	(s9)
		edge   node [left]  {?wd}  	(s7)
		(s4) edge   node [left]{?acc}    (s5)	 
		(s5) edge  [bend left = 30] node {?amo}    (s6)	 
		(s6) edge   [bend left = 60] node {!rec}    (s0)	
		edge  [bend left = 30]  node {!ins}    (s5)
		(s7) edge   node {?amo}    (s8)	
		(s8) edge  [bend left = 30] node [below] {!mon} (s0)
		edge  [bend left = 30] node {!ins}   (s7)
		(s9) edge   node [below] {!sho} (s0);
	\end{tikzpicture}	
	\caption{ATM specification $\mathcal{B}$}
	%\label{fig:spec-nconf-atm}
	\label{fig:spec-n-conf-atm}
\end{figure}
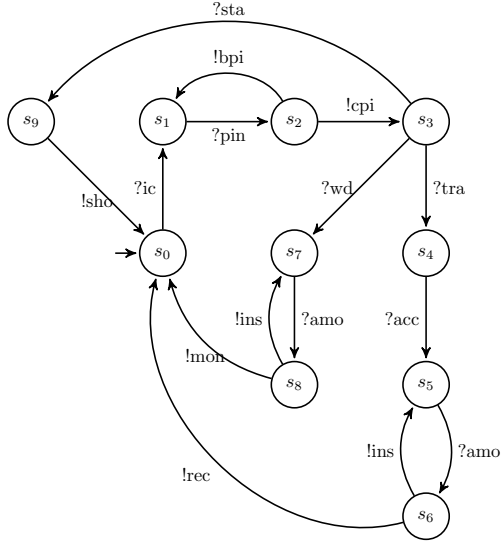

An withdrawal operation is then specified by the IOLTS $\mathcal{A}$ of Figure~\ref{fig:spec-atm1}. 
Note that if the requested amount (amo) is greater than the available amount (ins) then the withdrawal cannot be performed and the process reaches state $s_3$ where a new withdrawal operation can be selected again.
% !TeX spellcheck=en_US
We also specify  some additional functionalities on the IOLTS $\mathcal{B}$ of Figure~\ref{fig:spec-n-conf-atm}. 
In this case we consider not only the withdrawal (wd) operation but also a transfer (tra) and a statement (sta).
% !TeX spellcheck=en_US

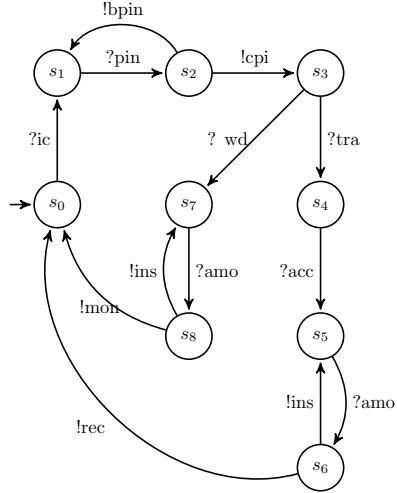
\begin{figure}[bht]
	\center
	\begin{tikzpicture}[->,>=stealth',shorten >=1pt,auto,node distance=2.5cm, semithick,initial text =,scale=0.7,transform shape]
		
		\node[initial, state]       (s0) {$s_0$};
		\node[state]                (s1) [above of=s0]{$s_1$};
		\node[state]                (s2) [right of=s1] {$s_2$};
		\node[state]               	(s3) [right of=s2]{$s_3$};
		\node[state]       			(s4) [below of=s3]{$s_4$};
		\node[state]                (s5) [below of=s4]{$s_5$};
		\node[state]                (s6) [below of=s5] {$s_6$};
		\node[state]               	(s7) [below of=s2]{$s_7$};
		\node[state]               	(s8) [below of=s7]{$s_8$};
		
		\path 
		(s0) edge               node {?ic}    (s1)	
		(s1) edge               node {?pin}   	(s2)
		(s2) edge               node {!cpi}   	(s3)
		edge  [bend right = 60] node [above]{!bpin}   	(s1)   
		(s3) edge   node {?tra}   	(s4)
		edge   node [left]  {? wd}  	(s7)
		(s4) edge   node [left]{?acc}    (s5)			 
		(s5) edge  [bend left = 30]  node  {?amo}    (s6) 
		(s6) edge   [bend left = 60] node {!rec}    (s0)
		edge   node {!ins}    (s5)	
		(s7) edge   node {?amo}    (s8)	
		(s8) edge  [bend left = 30] node [below] {!mon} (s0)
		edge  [bend left = 30] node {!ins}   (s7);
	\end{tikzpicture}	
	\caption{IUT $\mathcal{Z}$}
	%\label{fig:iut-atm}
	\label{fig:iut-atm2}
\end{figure}

Assume the IOLTS $\mathcal{Z}$ depicted at Figure~\ref{fig:iut-atm2} as an IUT that implements the withdrawal (wd) and transfer (tra) operations. 
We observe that if the requested  amount (amo) in a withdrawal is greater than the available amount the IUT model reaches state $s_7$ where the user can choose a new amount. 
% !TeX spellcheck=en_US

Now as a first testing scenario we check whether the IUT $\mathcal{Z}$  conforms to the specification $\mathcal{A}$. 
JTorx and \everest have returned the same verdict of conformance when running the \ioco checking relation. 
By contrast the language-based conformance that has been implemented in \everest tool could detect a fault. 
We considered the set of desirable behaviors $D = \{\text{ic pin cpi wd amo ins amo}\}$, \emph{i.e.}, a sequence of actions where the account balance is not enough according to the requested withdrawal, and the user must provide a new value. 
\everest has generated the test case $\{ic \xrightarrow[]{} pin \xrightarrow[]{} cpi \xrightarrow[]{} wd \xrightarrow[]{} amo \xrightarrow[]{} ins \xrightarrow[]{} amo\}$ since the behavior given by $D$ is not observable in the specification but it is implemented on the IUT $\mathcal{Z}$. 

In a second scenario we aim to check the reliability over the verdict obtained by JTorx tool using \ioco conformance, since it modifies the original underspecified models (See Section~\ref{sec:comparative}). 
First, \textit{self-loop} transitions are added at underspecified states in order to obtain an \textit{input-enabled} model. 
Since the IUT $\mathcal{Z}$ is underspecified, JTorx must guarantee that all states on the IUT are \textit{input-enabled}, before checking whether $\mathcal{Z}$ \ioco conforms to the specification $\mathcal{B}$. 
Therefore, the original behavior of the IUT is modified and now a fault can be detected by the test case $\{ic \xrightarrow{} pin \xrightarrow{} cpin \xrightarrow{} sta\}$. 

We have also applied this second scenario to the \everest using the \ioco relation. 
At this time any fault was detected since the fault behavior $\{ic \xrightarrow{} pin \xrightarrow{} cpin \xrightarrow{} sta\}$ was not specified in the IUT $\mathcal{Z}$. 
We see that the detection of this fault by JTorx is actually a false positive, since it appears due to an extra behavior added by the tool once the IUT $\mathcal{Z}$  is modified to become an \textit{input-enabled} model. 

We remark that \everest could have detected this same fault when checking \ioco conformance over the same modified model. 
We note that $ic$ is the unique action that is defined at the initial state $s_0$ of IUT $\mathcal{Z}$. 
When JTorx turns $\mathcal{Z}$ into an \textit{input-enabled} model 
all input actions become enabled at all states, which is inconsistent with the real functionality. 
For instance, we see that the action $amo$, \emph{i.e.}, the amount value to be withdrawn, is now defined at state $s_0$. 
However, if a transfer operation ($tra$) is chosen instead of a withdrawal ($wd$), the amount value to be withdrawn should not be enabled at this moment. 
Notice those changes performed over the underspecified models have changed the original behavior of the IUT, leading to an inaccurate conformance checking result w.r.t. the real functionality of the ATM. 

\begin{figure}[hbt]
	\center
	\begin{tikzpicture}[->,>=stealth',shorten >=1pt,auto,node distance=2.5cm, semithick,initial text =,scale=0.7,transform shape]
		
		\node[initial, state]       (s0) {$s_0$};
		\node[state]                (s1) [right of=s0]{$s_1$};
		\node[state]                (s2) [right of=s1] {$s_2$};
		\node[state]               	(s3) [below of=s2]{$s_3$};	
		\node[state]               	(s7) [below of=s3]{$s_4$};
		\node[state]               	(s8) [below of=s1]{$s_5$};

		\path 
		(s0) edge               node {?ic}    (s1)	
		(s1) edge               node {?pin}   	(s2)
		(s2) edge               node {!cpi}   	(s3)
		edge  [bend right = 60]  node [above]{!bpi}   	(s1)
		(s3) edge   node {?wd}   	(s7)		 	
		(s7) edge   node {!mon}    (s8)	
		(s8) edge  [bend left = 30] node [below] {!mon} (s0)
		edge  [bend left = 30] node {!ins}   (s3);
	\end{tikzpicture}	
	\caption{IUT $\mathcal{Y}$}
	\label{fig:iut-atm3}
\end{figure}
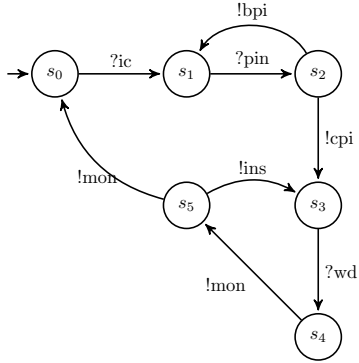

Next consider the IOLTS $\mathcal{Y}$ depicted in Figure~\ref{fig:iut-atm3} as a new IUT. 
% !TeX spellcheck=en_US
Notice that $\mathcal{Y}$ differs from the specification 
$\mathcal{A}$ depicted in Figure~\ref{fig:spec-atm1} over the transitions $(s_4,?amo,s_5)$ and $(s_4,!mon,s_5)$, respectively. 
So we can see that the IUT allows a withdrawal operation with no checking over the balance ($amo$) before releasing the money ($mon$). 
By contrast the balance is checked before releasing the money on the specification if the balance on account is positive. 
We set up the fault model by a bound of 6 states on the IUT models and \everest has generated 80 TPs based on the specification $\mathcal{A}$. 
After we have run the test suite over the IUT $\mathcal{Y}$, a fault verdict was declared by our tool by means of the path $ic \xrightarrow{} pin \xrightarrow{} cpin \xrightarrow{} wd$.
For the sake of completeness we also applied this last scenario to JTorx tool, and a fault was detected by the test case $?ic, \delta, ?pin, !bpi, ?pin, !cpi, ?wd, !mon$.

% !TeX spellcheck=en_US
\subsection{A Comparative Analysis} \label{sec:comparative}

Here we list some main aspects and compare \everest and JTorx tools. 
We have seen that both tools provide a mechanism of  test generation, test run and \ioco conformance checking. 
However notice that \everest also provides the more general conformance checking based on regular languages~\cite{bonifacio2018}. 
Further, \everest allows a complete test generation not only for the \ioco relation but also for this more general conformance relation with a wider range of possibilities to specify desirable and undesirable behaviors. 

In the test generation process of JTorx an exhaustive strategy is employed to generate test cases. 
This exhaustive process renders to an exponential state space when exploring the model which is unfeasible in practice. 
In other direction \everest is more flexible and allows a complete test suite generation by setting the maximum number of states over the IUTs to be considered in a fault model. 

JTorx  implements a random approach when choosing transitions to induce paths over the specification to generate test suites. 
\everest, however, only applies a random approach over the language-based conformance relation when desirable and/or undesirable behaviors are not provided by the tester. 
In this case, the test run is reduced to the problem of checking isomorphism between the IUT and the specification model. 

We also note that both tools implement an online testing approach 
%to generate and run test suites 
when IUTs are provided together with the specification model. 
However only \everest provides an offline test generation process using the notion of multigraph and test purposes. 

In the conformance checking process, JTorx defines an online strategy, where a test case is generated and after that is applied over the IUT. 
\everest follows an offline process where a test suite is generated and then applied to the IUT. 
However we remark that \everest also has an online alternative process when checking conformance where each test case extracted from the fault model is already applied to IUT in sequel. 
Table~\ref{tab:analise-comp-ferram} summarizes these aspects. 
\begin{table}[htb]
%	\vspace*{-3ex}
		\centering
	\begin{minipage}[t]{.5\textwidth}
		{\fontsize{9}{10}\selectfont
		\caption{\label{tab:analise-comp-ferram}Methods and Features}
		\begin{tabular}{|l|c|c|}
			%			\caption{Comparative analysis: methods.}
			\hline
			&\multicolumn{1}{l|}{JTorx} & \cellcolor{gray!50}\textbf{ Everest} \\ \hline
			\cellcolor{gray!25}\begin{tabular}[c]{@{}l@{}}\textbf{Conformance checking} \end{tabular}
			&\cellcolor{gray!25}&\cellcolor{gray!25}\\
			\ioco theory  & X   &  \cellcolor{gray!50}\textbf{X}  \\ 
			Language-based  & - & \cellcolor{gray!50}\textbf{X} \\ 
			
			\cellcolor{gray!25}\textbf{Generation} &\cellcolor{gray!25}&\cellcolor{gray!25}\\
			Test suite generation    & X  & \cellcolor{gray!50}\textbf{X} \\ 
			
			\cellcolor{gray!25}\textbf{Test strategy} &\cellcolor{gray!25}&\cellcolor{gray!25}\\
			online/offline   & X   &  \cellcolor{gray!50}\textbf{X}\\ 
%			offline   & X  & \cellcolor{gray!50}\textbf{X} \\ 				
			Test purpose   & X  &  \cellcolor{gray!50}\textbf{X}  \\ 
			Random approach  & X &  \cellcolor{gray!50}\textbf{X}  \\ 
%			\cellcolor{gray!25}\textbf{Implementation} &\cellcolor{gray!25}&\cellcolor{gray!25}\\
%			Language  &  Java & \cellcolor{gray!50}\textbf{Java} \\ 
			\hline
		\end{tabular}
	}
		\end{minipage}%
\end{table}

We also probe some properties over the specification and IUT models, test verdicts and strategies of testing. 
See Table~\ref{tab:analise-comp-ferra2}. 
\begin{table}[htb]
				\centering
	\begin{minipage}[t]{0.5\textwidth}
		{\fontsize{9}{10}\selectfont
			\caption{\label{tab:analise-comp-ferra2} Properties and Tools}
			\begin{tabular}{|l|c|c|}
				%			\caption{Comparative analysis: properties.}
				\hline
				&  \multicolumn{1}{l|}{JTorx} & \cellcolor{gray!50}\textbf{Everest} \\ \hline
				\cellcolor{gray!25}\begin{tabular}[c]{@{}l@{}}\textbf{Properties} \end{tabular}
				&\cellcolor{gray!25}&\cellcolor{gray!25}\\
				\begin{tabular}[c]{@{}l@{}}Underspecified models\end{tabular}   
				& X  & \cellcolor{gray!50}\textbf{X} \\ 
				\begin{tabular}[c]{@{}l@{}}Require input-enabledness \end{tabular} 
				& X  & \cellcolor{gray!50}\textbf{-} \\ 
				\begin{tabular}[c]{@{}l@{}}Quiescence\end{tabular} & X & \cellcolor{gray!50}\textbf{X} \\ 
				\cellcolor{gray!25}\textbf{Veredicts} &\cellcolor{gray!25}&\cellcolor{gray!25}\\
				Test run &    X %\footnote{$\{pass, fail, inconclusive\}$}  
				&  \cellcolor{gray!50}\textbf{X} \\ 
				Conformance &   X & 
				%\footnote{conform/non-conform} &  
				\cellcolor{gray!50}\textbf{X} \\ 		
				\cellcolor{gray!25}\textbf{Test mode} &\cellcolor{gray!25}&\cellcolor{gray!25}\\
				White/black boxes testing   & X  & \cellcolor{gray!50}\textbf{X}  \\ 
				%Black-box testing   & X & \cellcolor{gray!50}\textbf{X} \\ 
				\hline
			\end{tabular}
		}
	\end{minipage}
	%	\vspace*{-3ex}
\end{table}
The \ioco conformance checking naturally imposes some restrictions over the models. 
Underspecified models, for instance, are not allowed on IUT side and 
their internal structure must be changed to guarantee the input-enabledness.

The language-based conformance relation does not require any restriction, that is, the more general method can deal with underspecified specification and implementation models. 
So \everest can handle underspecified models when checking conformance and generating test suites with no change over the models. 
JTorx, on the other hand, must explore the entirely structure of 
the models to add new transitions in order to guarantee input-enabledness.

%Another important issue on testing reactive systems is quiescence. 
%An IOLTS model is quiescent if there exists at least one state where no output is produced~\cite{bonifacio2018,tretmans2008}. 
Both tools also deal with quiescence by adding self-loops with $\delta$ actions at quiescent states and  
%Both tools can also 
give verdicts when checking conformance and running test cases with small variations. 
%Verdicts of test runs may be $\{pass, fail, inconclusive\}$:  \textit{pass} indicates that the IUT complies with the specification; \textit{fail} 
%reveals a fault when the IUT does not conform to the specification; and 
%an inconclusive verdict says any fault could be detected applying a specific TP.  

% !TeX spellcheck=en_US
\section{Practical Evaluation: Experiments}\label{sec:experiments}

We have also run some practical experiments to evaluate the tools' performance. 
First we provide some experiments to compare the conformance verification process of both tools. 
In addition we assay \everest when generating and running test suites using test purposes. 
All experiments are organized by means of Research Questions (RQs) in order to achieve a goal for each group of scenarios. 
We have performed these experiments on Intel Core i5 1.8 GHz CPU, 8 GB of RAM memory on Windows 10 operating system. 

%Cada teste foi executado uma única vez, devido a quantidade enorme de experimentos distintos realizados e o tempo total consumido. 
%No total foram realizados 35.640 experimentos, entre verificações de conformidade e execuções de TPs nas IUTs, e 740 multigrafos gerados.

\subsection{Conformance checking: \everest and JTorx}

We first evaluate several parameters related to the specification and IUT  models, such as the number of states and the number of input/output actions. 
Despite of these parameters we also consider scenarios with verdicts of conformance and non-conformance. 
Notice that we have generated only input-enabled and deterministic models due to restrictions imposed by JTorx on checking conformance.
Further, in order to keep an unbiased analysis over the results all transitions were randomly generated to construct the models, but in such way that all required properties were still satisfied. 

All IUT models  ioco-conform to their respective specifications were obtained as submachines while IUTs non-ioco-conform were constructed by changing transitions based on their specifications according to a certain percentage of modification. 
Regarding IUTs with more states than the number of states on specifications, \emph{i.e.} $m>n$, we added new states and transitions to the IUTs.

Here each experiment is defined by a conformance checking between a specification and ten different IUTs. 
So a group of experiments with ten specifications and ten IUTs for each specification results in hundred runs. 
The processing time that is shown off in the graphics corresponds to the average processing time for all experiments in a group. 

%Uma das dificuldades encontradas para se medir o tempo, de execução das verificações de conformidade, na ferramenta JTorx foi que,  apesar da ferramenta ter código aberto, os fontes não estavam todos disponíveis, não sendo possível realizar as modificações necessárias para que o tempo de execução fosse registrado. 
%Portanto, para a medição do tempo de execução, das verificações de conformidade, foi desenvolvido um aplicativo de chamada para as ferramentas,  passando os parâmetros necessários, e então medindo o tempo total gasto no processo de verificação até  a apresentação do veredito. 
%A Seção~\ref{sec:ameacas-validade} apresenta as ameaças à validade dos experimentos realizados. 

%Everest has been modified to generate the same test suite as generated by JTorx, defining the desired number of test cases. 
%Again, we have introduced such modification in order to gauge  unbiased over the  experiments when comparing both tools.

% variação i-o
\subsubsection{Varying the size of alphabets}

In this first scenario we investigate the impact on checking conformance when varying the number of input/output actions. 
So the \textbf{RQ}, in this case, is:
\textit{``Does the size of the input and output alphabets impact the processing time on checking conformance?"}. 
In order to answer this question we have run experiments over specifications with 10 states and IUTs with 15, 25 and 35 states, for both verdicts of conformance and non-conformance. 
We consider IOLTS models with 12 symbols: one group with 2 inputs and 10 outputs; and a second group with 10 inputs and 2 outputs.

\begin{figure}[hbt]
	%	\vspace*{-1ex}
	\begin{center}
		\begin{tabular}{@{}cc@{}}
			\subfloat[][2 inputs, 10 outputs, ioco
			\label{fig:io-ioco-conf-2-10}]{
				\includegraphics[scale=0.3, trim=.5cm 10cm 13cm .5cm,clip]{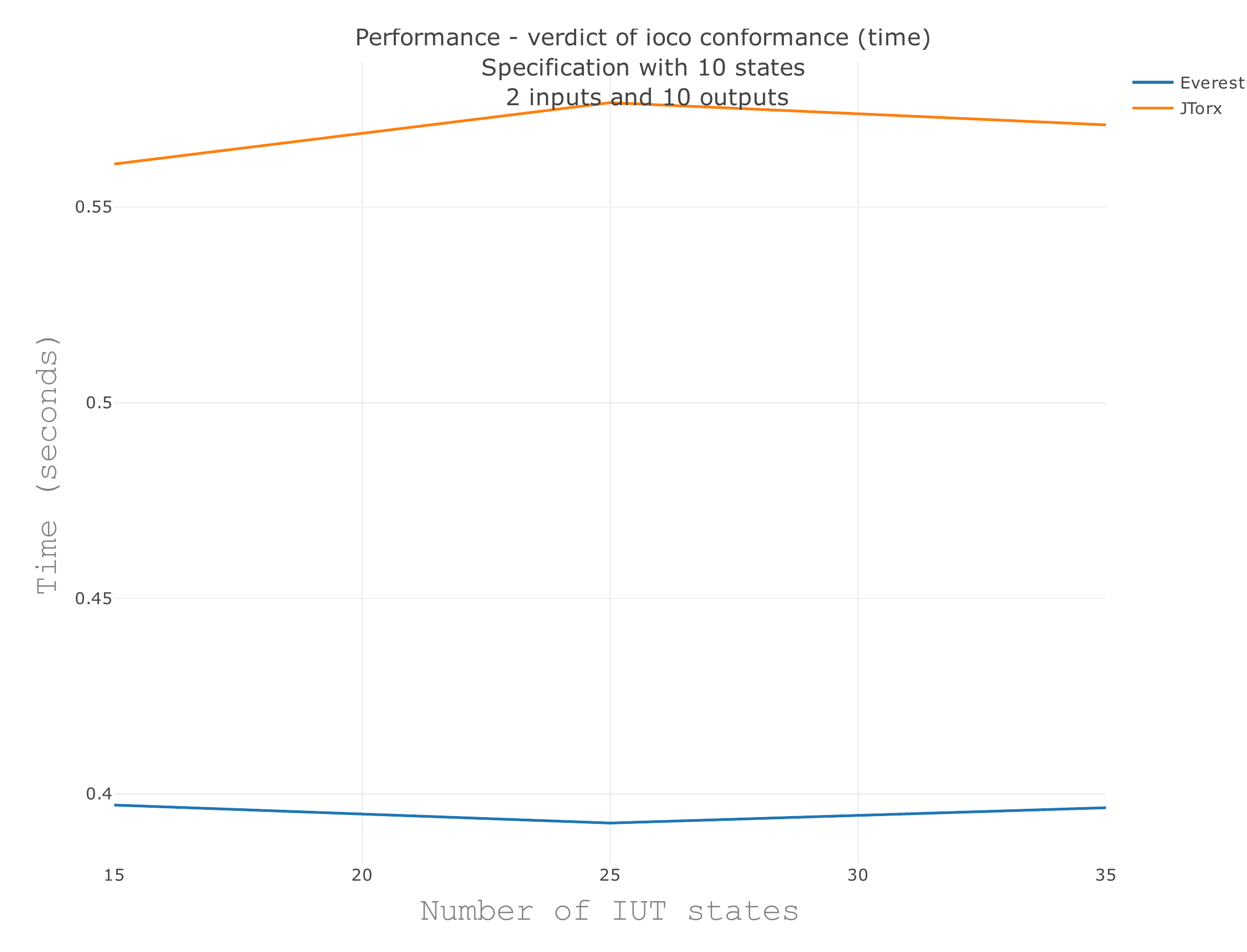}	
			}
			&
			\subfloat[10 inputs, 2 outputs, ioco
			\label{fig:io-ioco-conf-10-2}]{
				\includegraphics[scale=0.3, trim=.5cm 10cm 13cm .5cm,clip]{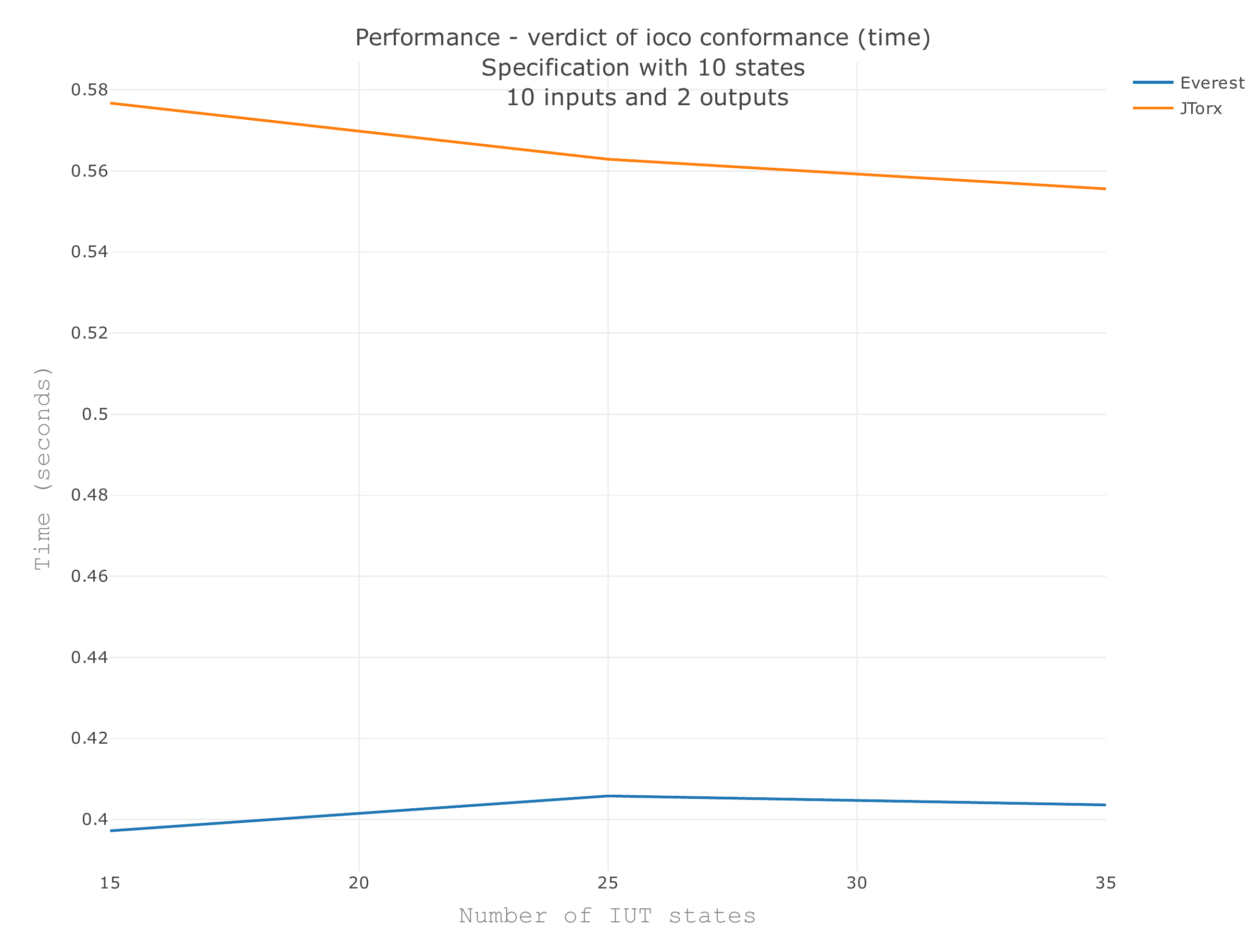}
			}			
		\end{tabular}
	\end{center}
	%	\vspace*{-1ex}
	\caption{Varying I/O alphabets and conformance \label{fig:experimento-ioco-conf} }
	%	\vspace*{-3ex}
\end{figure}

We observe at Figure~\ref{fig:experimento-ioco-conf} that the impact is not expressive for experiments with verdicts of conformance. 
We notice that our tool is $2,56\%$ faster when running models with 2 inputs and 10 outputs than when checking models whose alphabets have 10 inputs and 2 outputs, respectively. 
In the opposite direction, JTorx is approximately $3,51\%$ faster over 
models with 10 inputs and 2 outputs than running over models with 2 inputs and 10 outputs.

\begin{figure}[hbt]
	%	\vspace*{-1ex}
	\begin{center}
		\begin{tabular}{@{}cc@{}}
			\subfloat[][2 inputs, 10 outputs, non-ioco
			\label{fig:io-ioco-n-conf-2-10}]{
				\includegraphics[scale=0.3, trim=.5cm 10cm 13cm .5cm,clip]{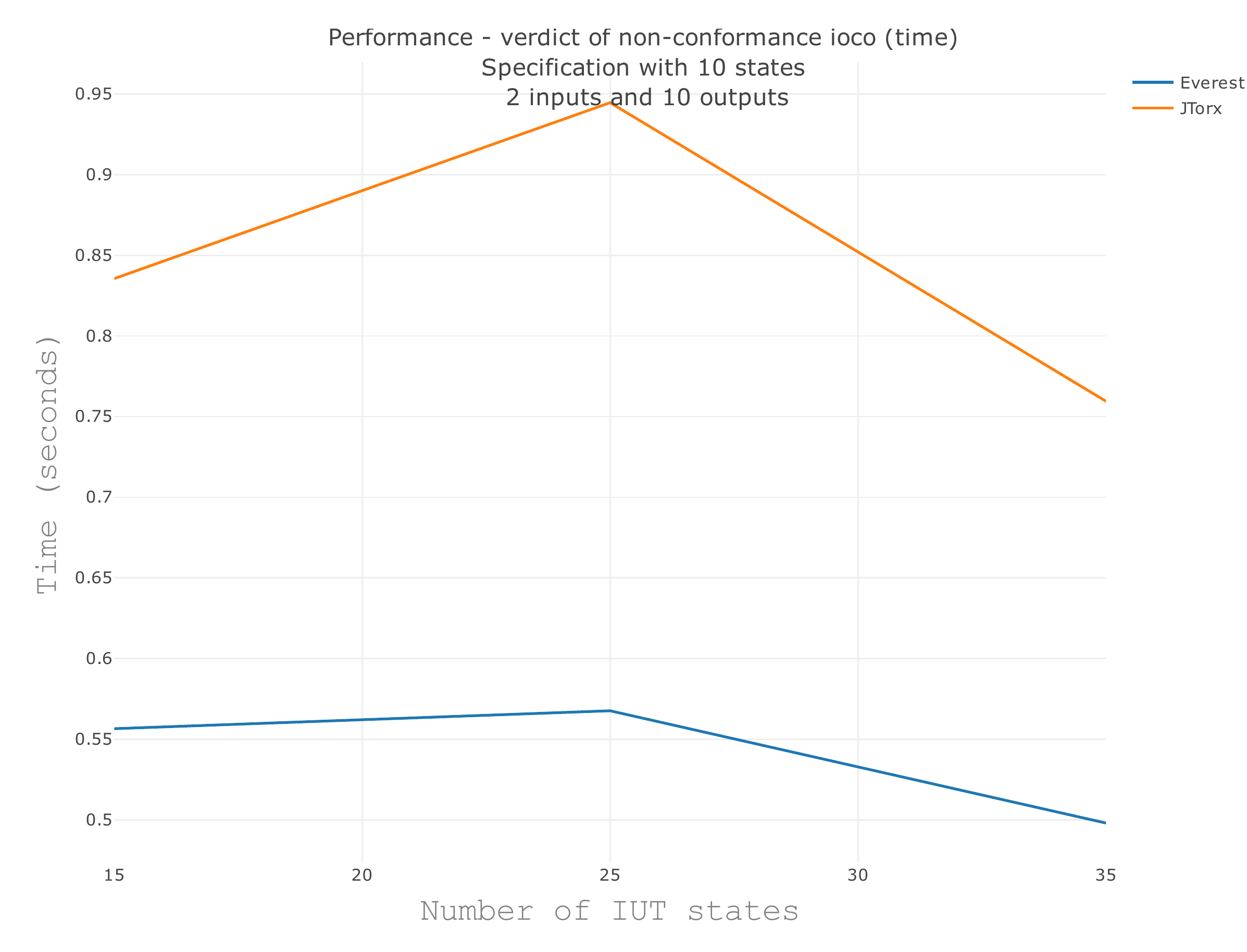}
			}
					&
			\subfloat[10 inputs, 2 outputs, non-ioco  
			\label{fig:io-ioco-n-conf-10-2}]{
				\includegraphics[scale=0.3, trim=.5cm 10cm 13cm .5cm,clip]{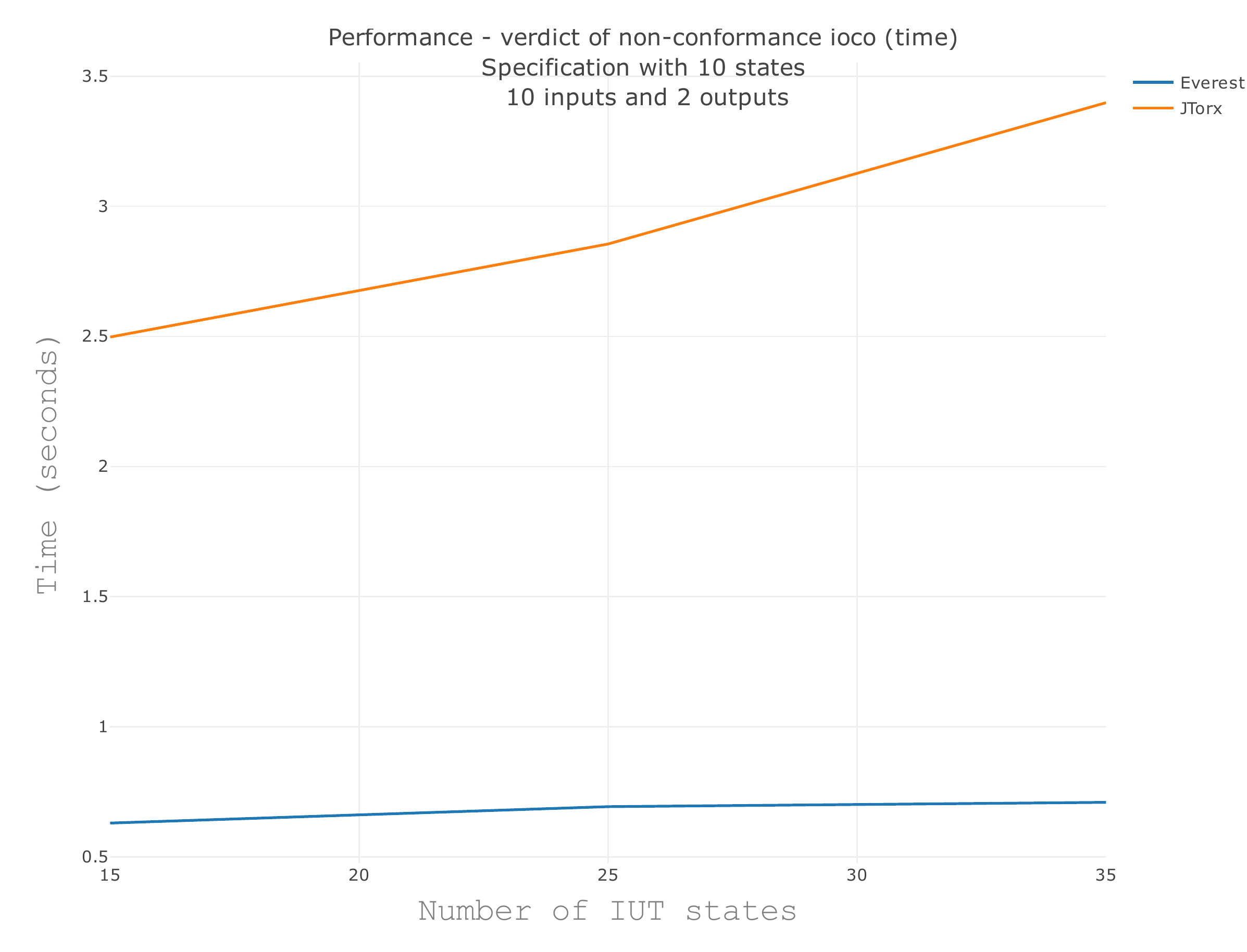}
			}
			
		\end{tabular}
	\end{center}
	%	\vspace*{-1ex}
	\caption{Varying I/O alphabets and non-conformance \label{fig:experimento-ioco-n-conf} }
	%	\vspace*{-3ex}
\end{figure}

In contrast, we see at Figure~\ref{fig:experimento-ioco-n-conf}  an expressive impact on the verification time when running experiments with verdicts of non-conformance. 
Both tools have taken less processing time for models with 2 inputs and 10 outputs.  
Everest is $12,73\%$ to $42,86\%$ faster for models with 2 inputs and 10 outputs compared to models with 10 inputs and 2 outputs, whereas 
JTorx is around $200\%$ to $352\%$ faster for the same group of models. 
Usually in a practical application we have a higher number of input actions to be specified in real-world systems. 
That is, input alphabets with a large number of actions can weigh down the performance of JTorx tool. 
%\begin{figure}[hbt]
%%	\vspace*{-1ex}
%	\begin{center}
%		\begin{tabular}{@{}cc@{}}
%			\subfloat[][2 inputs, 10 outputs, ioco
%			\label{fig:io-ioco-conf-2-10}]{
%				\includegraphics[scale=0.28, trim=.5cm 10cm 13cm .5cm,clip]{figs/experimentos/i-o/ioco-conf/2inp-10out/tempo.pdf}	
%			}\\
%		%&
%			\subfloat[10 inputs, 2 outputs, ioco
%			 \label{fig:io-ioco-conf-10-2}]{
%				\includegraphics[scale=0.2, trim=.5cm 10cm 13cm .5cm,clip]{figs/experimentos/i-o/ioco-conf/10inp-2out/tempo.pdf}
%			}\\
%			\subfloat[][2 inputs, 10 outputs, non-ioco
%			 \label{fig:io-ioco-n-conf-2-10}]{
%				\includegraphics[scale=0.2, trim=.5cm 10cm 13cm .5cm,clip]{figs/experimentos/i-o/ioco-n-conf/2inp-10out/tempo.pdf}
%			}\\	
%%			&
%			\subfloat[10 inputs, 2 outputs, non-ioco  
%			\label{fig:io-ioco-n-conf-10-2}]{
%				\includegraphics[scale=0.28, trim=.5cm 10cm 13cm .5cm,clip]{figs/experimentos/i-o/ioco-n-conf/10inp-2out/tempo.pdf}
%			}
%			
%		\end{tabular}
%	\end{center}
%%	\vspace*{-1ex}
%	\caption{Varying I/O alphabets. \label{fig:experimento-ioco-conf-e-n-conf-io} }
%%	\vspace*{-3ex}
%\end{figure}

%%%% stress test com estados

\subsubsection{Varying the number of states}

We also performed some experiments varying the number of states (and transitions)  to evaluate the tools' scalability. 
In this case the \textbf{RQ} is:
\textit{``How does the number of states in specifications and IUTs impact the processing time on checking conformance?"}. 
In order to answer this question we have run three groups of experiments: 
(i) specifications with 10 states and IUTs ranging from 20 to 200 states; 
(ii) specifications with 50 states and IUTs ranging from 60 to 200 states; and 
(iii) specifications with 100 states and IUTs varying from 110 to 200 states. 
We remark that all groups of IUTs were increased by 10 states in each group.

\begin{figure}[hbt]
	%	\vspace*{-1ex}
	\begin{center}
		\begin{tabular}{@{}cc@{}}
			\subfloat[][Specification with 10 states \label{fig:ioco-conf-10-}]{
				\includegraphics[scale=0.3, trim=.5cm 10cm 13cm .5cm,clip]{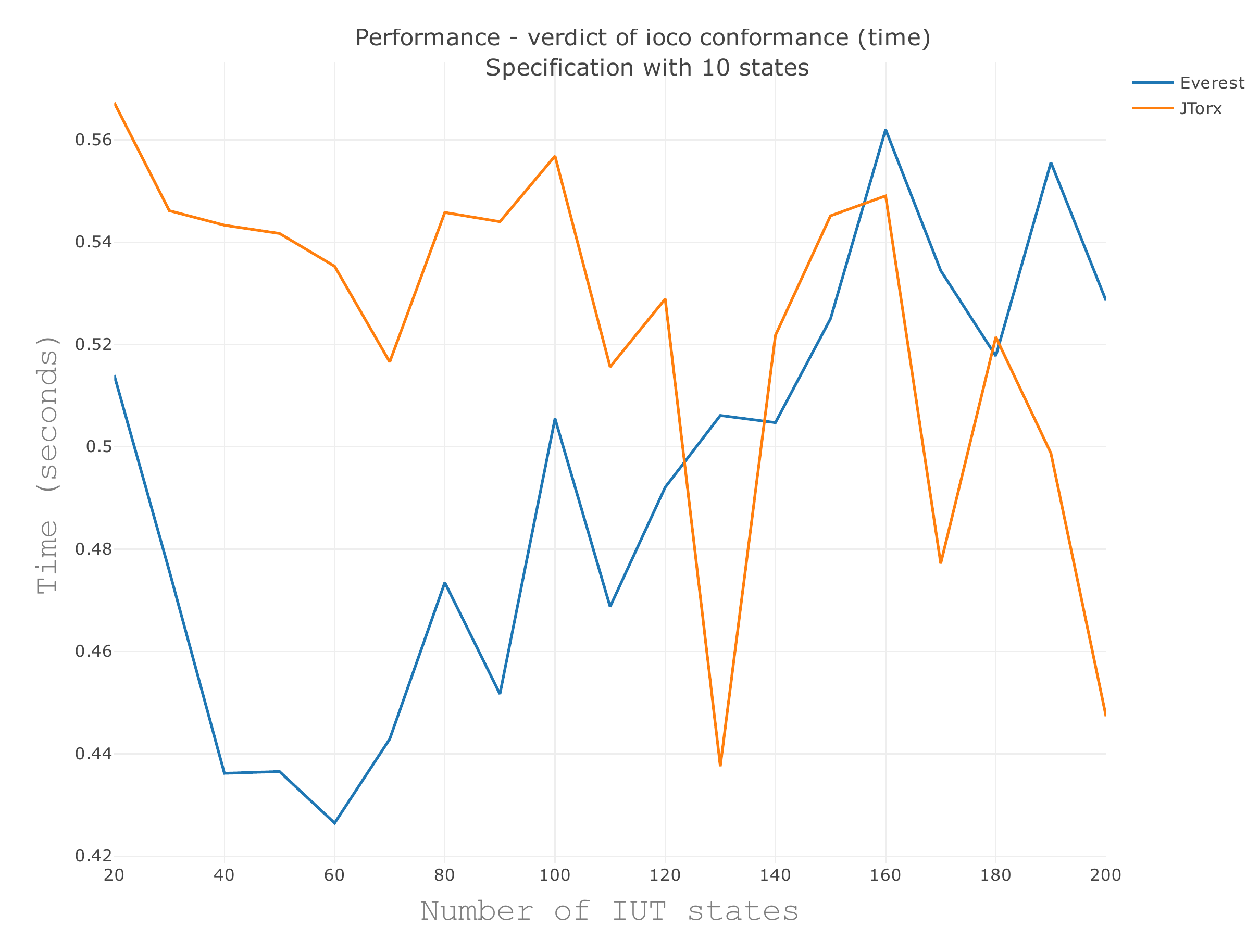}	
			}&
			\subfloat[Specification with 50 states \label{fig:ioco-conf-50}]{
				\includegraphics[scale=0.3, trim=.5cm 10cm 13cm .5cm,clip]{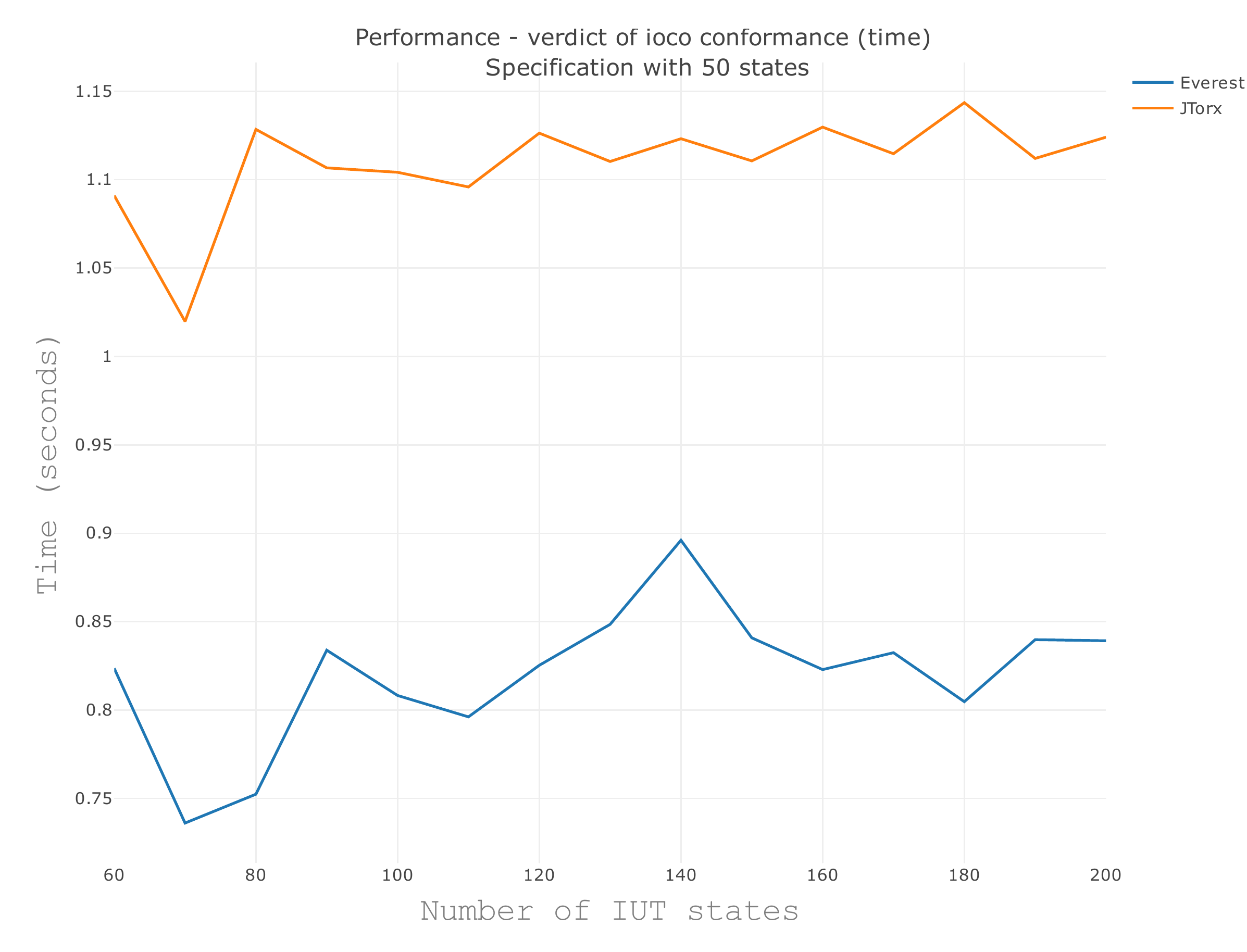}
			}\\ 			
		\subfloat[Specification with 100 states \label{fig:ioco-conf-100}]{
			\includegraphics[scale=0.3, trim=.5cm 10cm 13cm .5cm,clip]{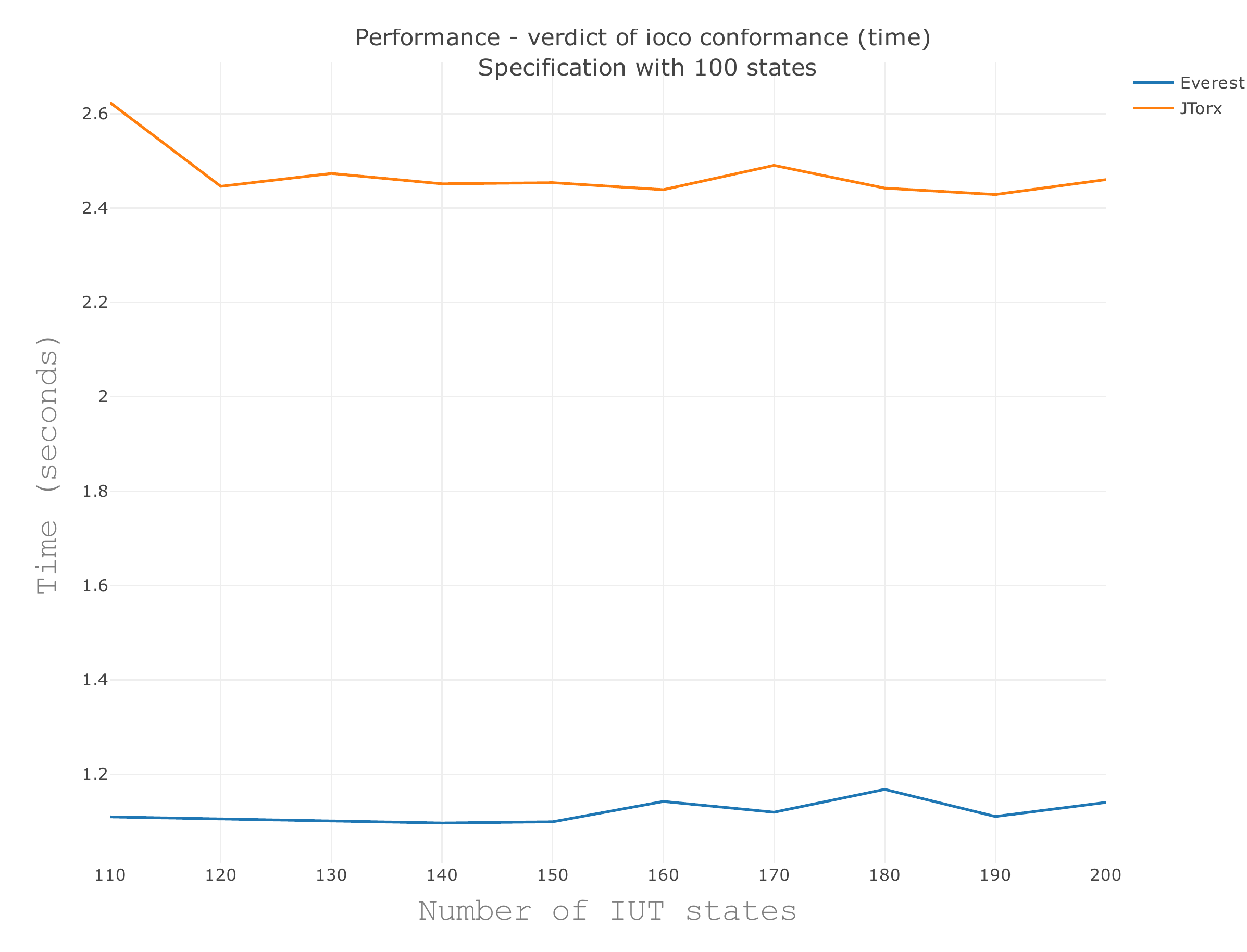}
		}
		\end{tabular}		
	\end{center}
	%	\vspace*{-2ex}
	\caption{Varying the number of states and conformance \label{fig:experimento-ioco-conf-10-50-100} }
	%	\vspace*{-3ex}
\end{figure}

Regarding experiments with conformance verdicts, specifications with 10 states and IUTs with up to 120 states, \everest reveals a better performance compared to JTorx. 
JTorx is just slightly better when considering IUTs with more than 120 states. 
See Figure~\ref{fig:ioco-conf-10-}. 

%\begin{figure}[hbt]
%%	\vspace*{-1ex}
%	\begin{center}
%		\begin{tabular}{@{}cc@{}}
%			\subfloat[][Specification with 10 states \label{fig:ioco-conf-10-}]{
%				\includegraphics[scale=0.2, trim=.5cm 10cm 13cm .5cm,clip]{figs/experimentos/stress/ioco-conf/10states/tempo.pdf}	
%			}
%			
%			&
%			\subfloat[Specification with 50 states \label{fig:ioco-conf-50}]{
%				\includegraphics[scale=0.2, trim=.5cm 10cm 13cm .5cm,clip]{figs/experimentos/stress/ioco-conf/50states/tempo.pdf}
%			}\\						
%		\end{tabular}
%		
%		\subfloat[Specification with 100 states \label{fig:ioco-conf-100}]{
%			\includegraphics[scale=0.2, trim=.5cm 10cm 13cm .5cm,clip]{figs/experimentos/stress/ioco-conf/100states/tempo.pdf}
%		}
%		
%	\end{center}
%%	\vspace*{-2ex}
%	\caption{Varying the number of states and ioco-conf. \label{fig:experimento-ioco-conf-10-50-100} }
%%	\vspace*{-3ex}
%\end{figure}
When running experiments with conformance verdicts, specifications with 50 and 100 states and groups of IUTs up to 200 states, \everest has always outperformed JTorx. 
See Figures~\ref{fig:ioco-conf-50} and~\ref{fig:ioco-conf-100}. 

\begin{figure}[hbt]
%		\vspace*{-1ex}
	\begin{center}
		\begin{tabular}{@{}cc@{}}
			\subfloat[][Specification with 10 states \label{fig:ioco-n-conf-10-}]{
				\includegraphics[scale=0.3, trim=.5cm 10cm 13cm .5cm,clip]{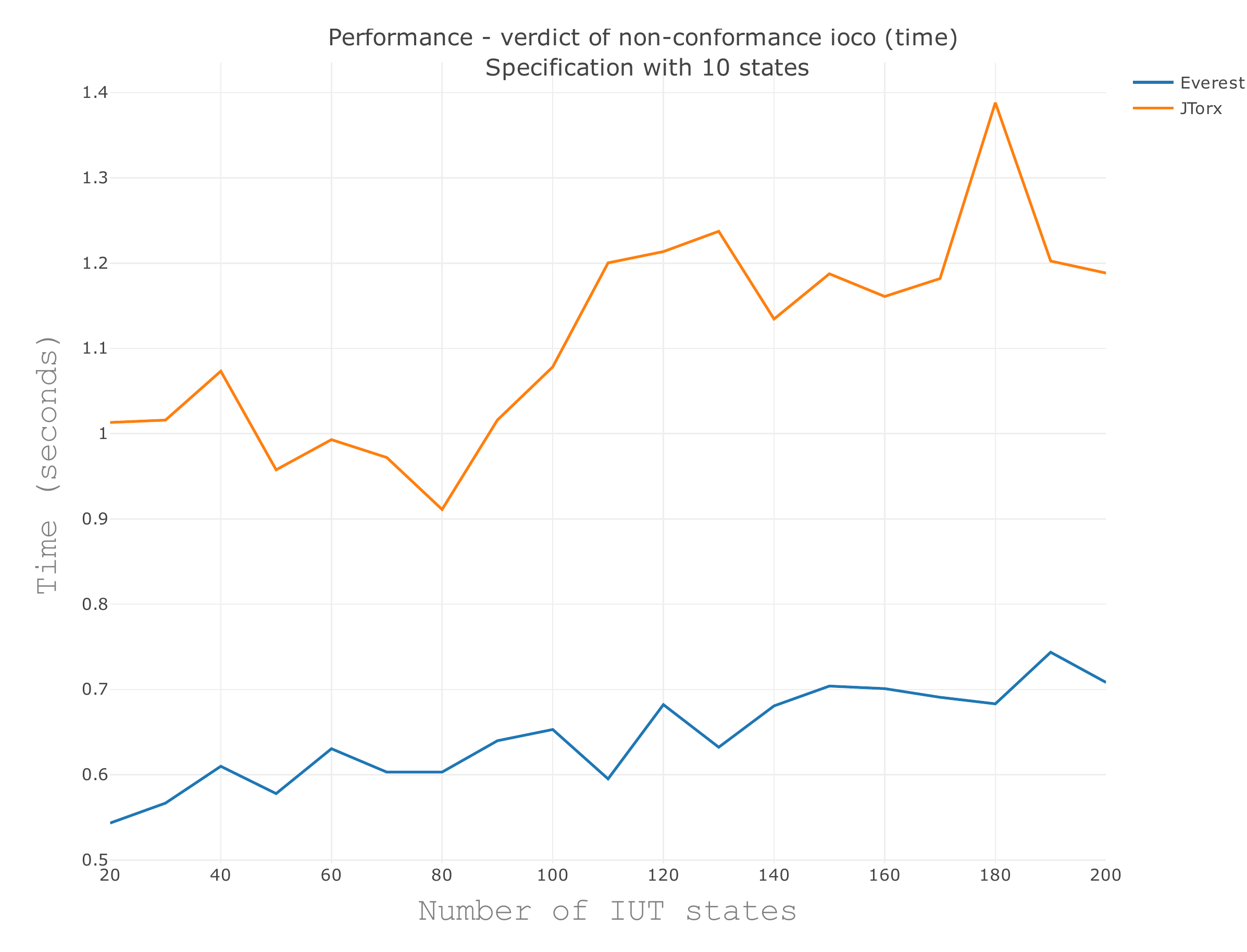}	
			} &
			\subfloat[Specification with 50 states \label{fig:ioco-n-conf-50}]{
				\includegraphics[scale=0.3, trim=.5cm 10cm 13cm .5cm,clip]{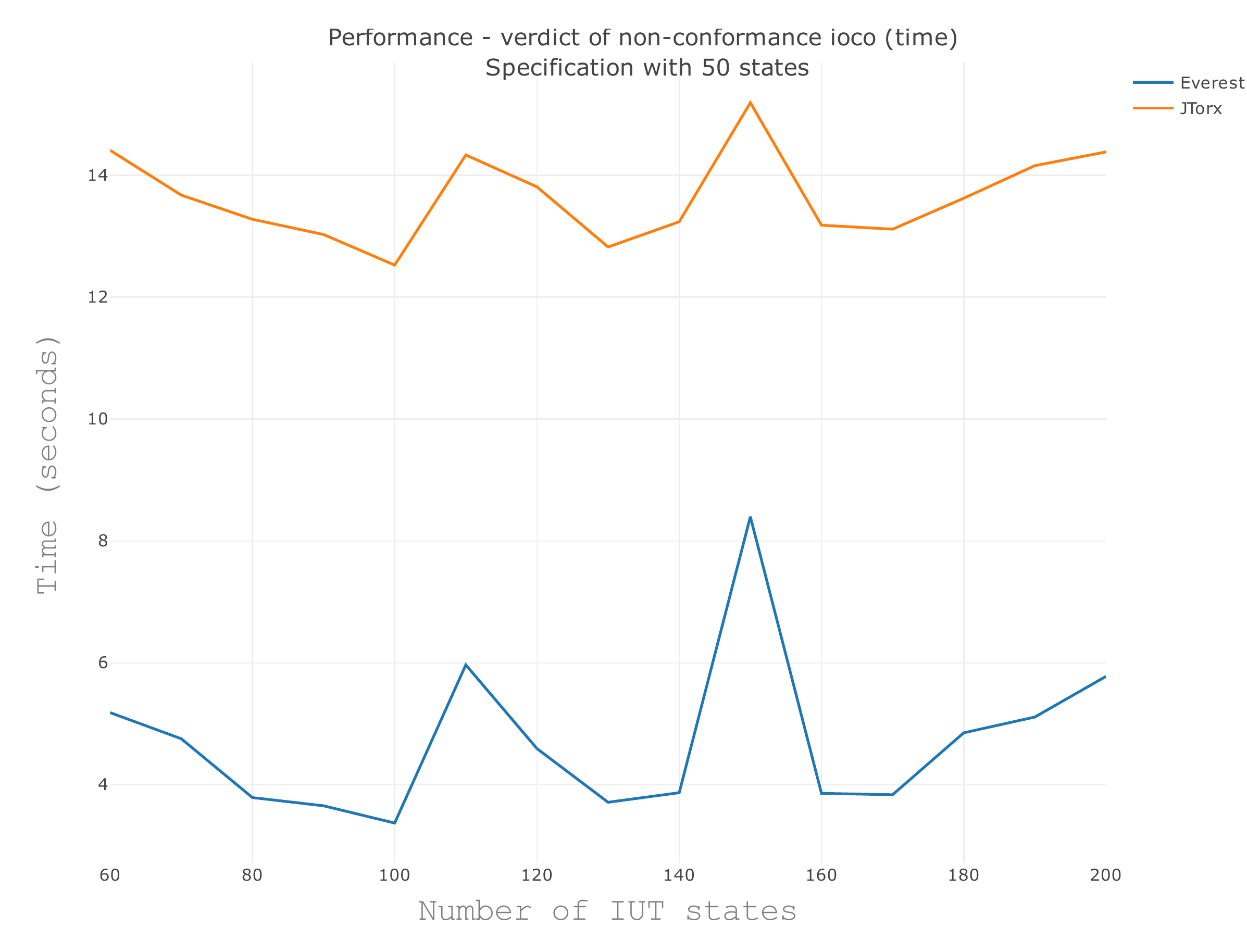}
			}\\								
		\subfloat[Specification with 100 states \label{fig:ioco-n-conf-100}]{
			\includegraphics[scale=0.3, trim=.5cm 10cm 13cm .5cm,clip]{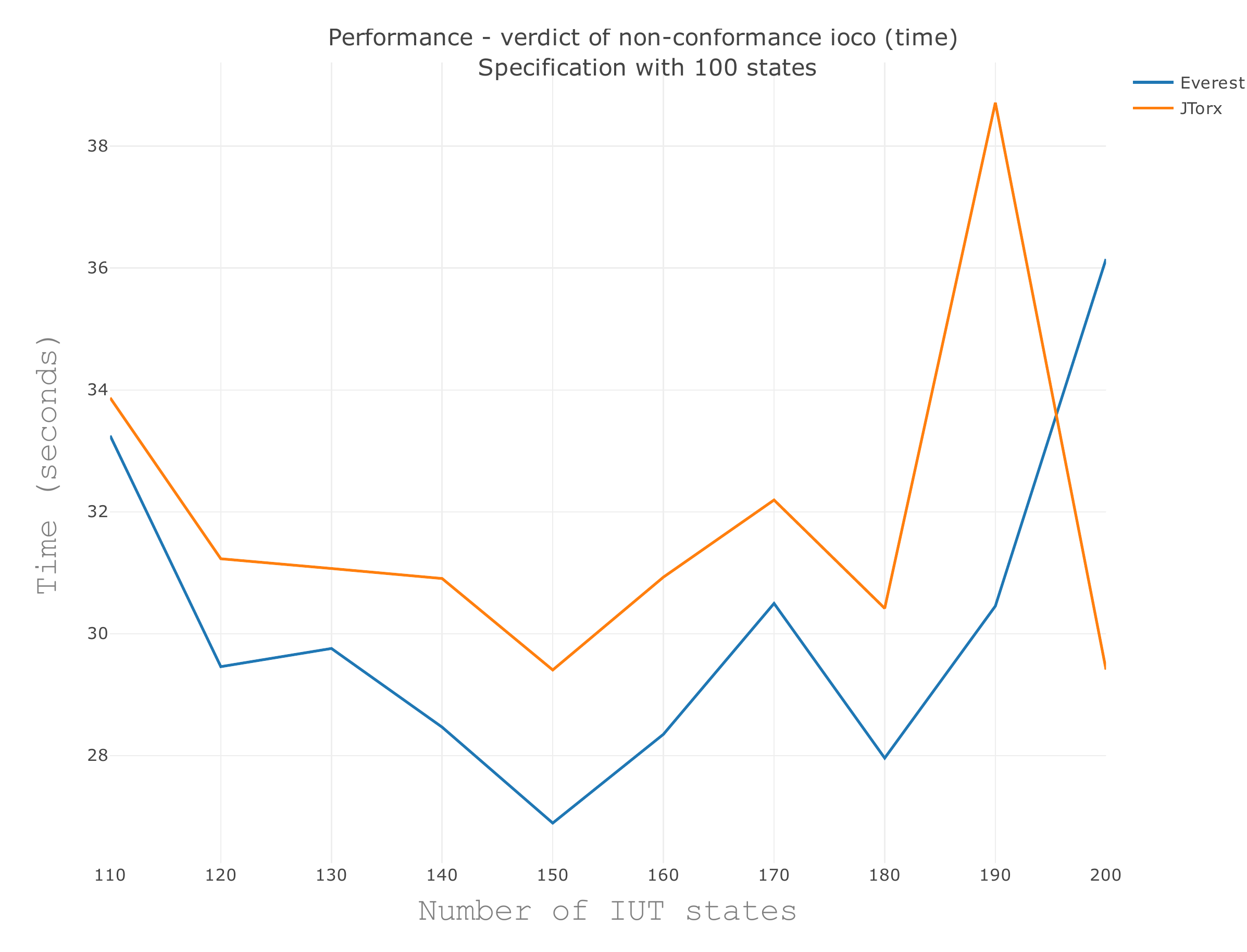}
		}
	\end{tabular}	
	\end{center}
%	\vspace*{-2ex}
	\caption{Varying the number of states and non-conformance \label{fig:experimento-ioco-n-conf-10-50-100} }
%		\vspace*{-3ex}
\end{figure}

Now considering experiments with non-conformance verdicts, specifications with 10 and 50 states, our tool always outperformed JTorx again for any group of IUTs. 
See Figures~\ref{fig:ioco-n-conf-10-} and~\ref{fig:ioco-n-conf-50}.  
JTorx tool has a better performance only  for IUTs with more than 200 states and specifications with 100 states.  
See Figure~\ref{fig:ioco-n-conf-100}. 

%%%%%%%%%%%%%%%%%%%%%%%%%%%%%%%%%%%%%%%%%%%%%%%%%%%%%%%%%%%%%%%%%%%%%%%%%%%%%%%%%%%%%%%%%%%%%%%%%%%%%%%%%%
% Geração & execução de teste
%%%%%%%%%%%%%%%%%%%%%%%%%%%%%%%%%%%%%%%%%%%%%%%%%%%%%%%%%%%%%%%%%%%%%%%%%%%%%%%%%%%%%%%%%%%%%%%%%%%%%%%%%%

%\subsection{Geração e Execução de Testes}

\subsection{\everest Test Suite Generation }

Now we evaluate our tool when running experiments of test suite generation using the more recent method~\cite{bonifacio2018}, where multigraphs are first constructed in order to then generate test purposes. 
Here we vary the number of states in the specification models and also the bound to be considered over the number of states on the IUTs. 
We also construct distinguishing IUTs from their respective specifications with a certain percentage of modification over the transitions in order to assess different scenarios. 

We remark that the test generation experiments were only performed using \everest for two main reasons:
(i)  an IUT must be also provided for the JTorx generation step since it implements an \textit{online} strategy; and % to generate and run test suites; 
(ii) a complete test suite cannot be constructed by JTorx since the process is finished at the very first  detected fault.
%(iii) dificuldade para capturar o tempo de geração e execução dos testes pelo JTorx, já que a ferramenta sobrepõe várias telas e processos, não permitindo uma avaliação de desempenho precisa.

\subsubsection{Multigraph generation step}

On generating multigraphs the associated \textbf{RQ} is: 
\textit{``What is the impact on the processing time when varying the number of states on specification models and the bound associated to the maximum number of states to be considered on IUT models?''}. 
To answer this question we consider specifications with 5 to 35 states and construct the respective multigraphs to get fault models for IUTs with 5 to 55 states. 
We increase the number of states by 10 for each group of IUTs and alphabets were fixed with 5 inputs and 5 outputs. 
Transitions were randomly generated to secure unbiasedness over the results.
Briefly the multigraph generation step takes into account the following scenarios:
(i) specifications with 5 states and \textit{m} from 5 to 55 states;
(ii) specifications with 15 states and \textit{m} from 15 to 55 states;
(iii) specifications with 25 states and \textit{m} from 25 to 55 states; and
(iv) specifications with 35 states and \textit{m} from 35 to 55 states.

\begin{figure}[hbt]
	%		\vspace*{-2ex}
	\begin{center}
		\begin{tabular}{@{}cc@{}}
			\subfloat[][Specifications with 5 and 15 states \label{fig:ger-mult-5-15}]{
				\includegraphics[scale=0.4]{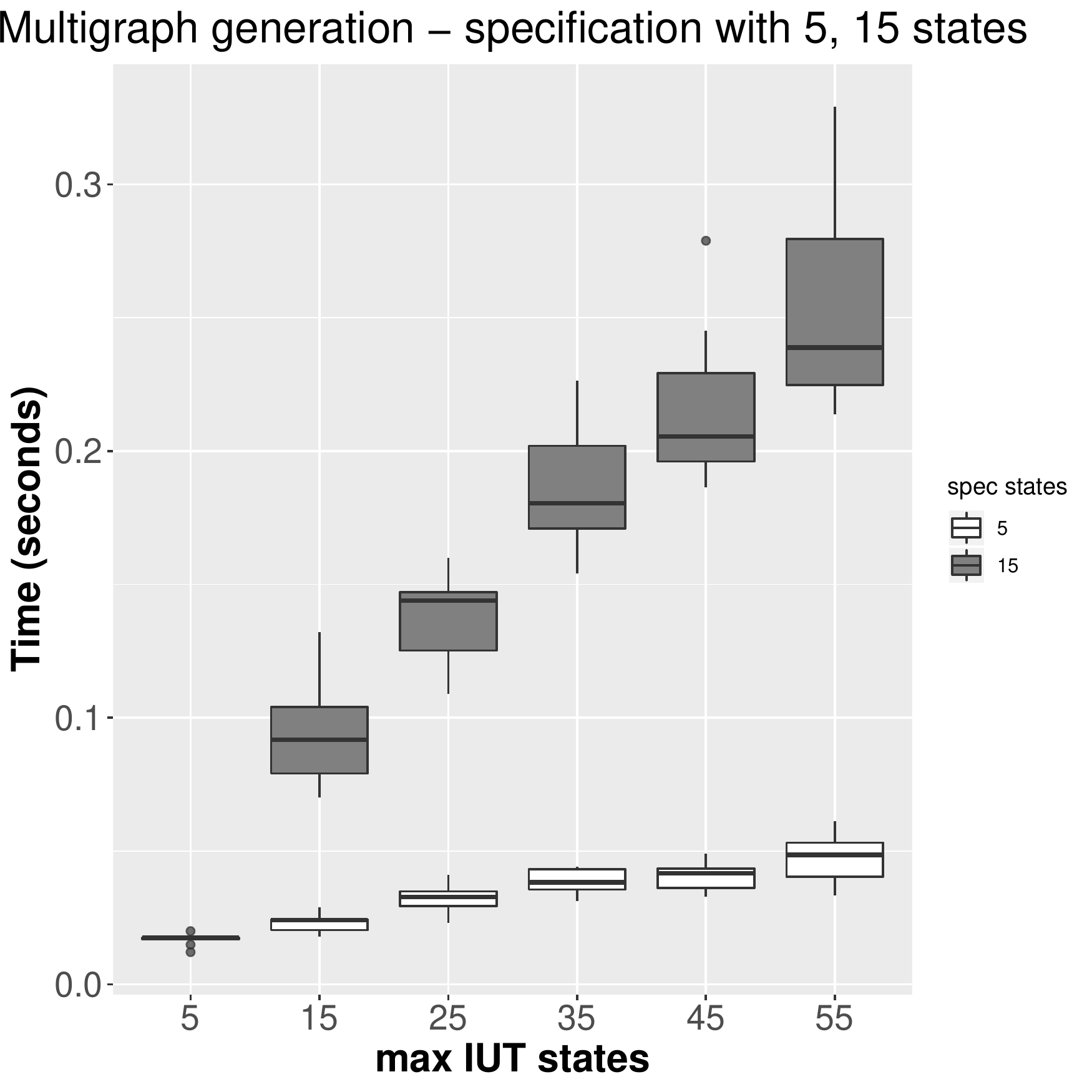}	
			} &
			\subfloat[Specifications with 25 and 35 states \label{fig:ger-mult-25-35}]{
				\includegraphics[scale=0.4]{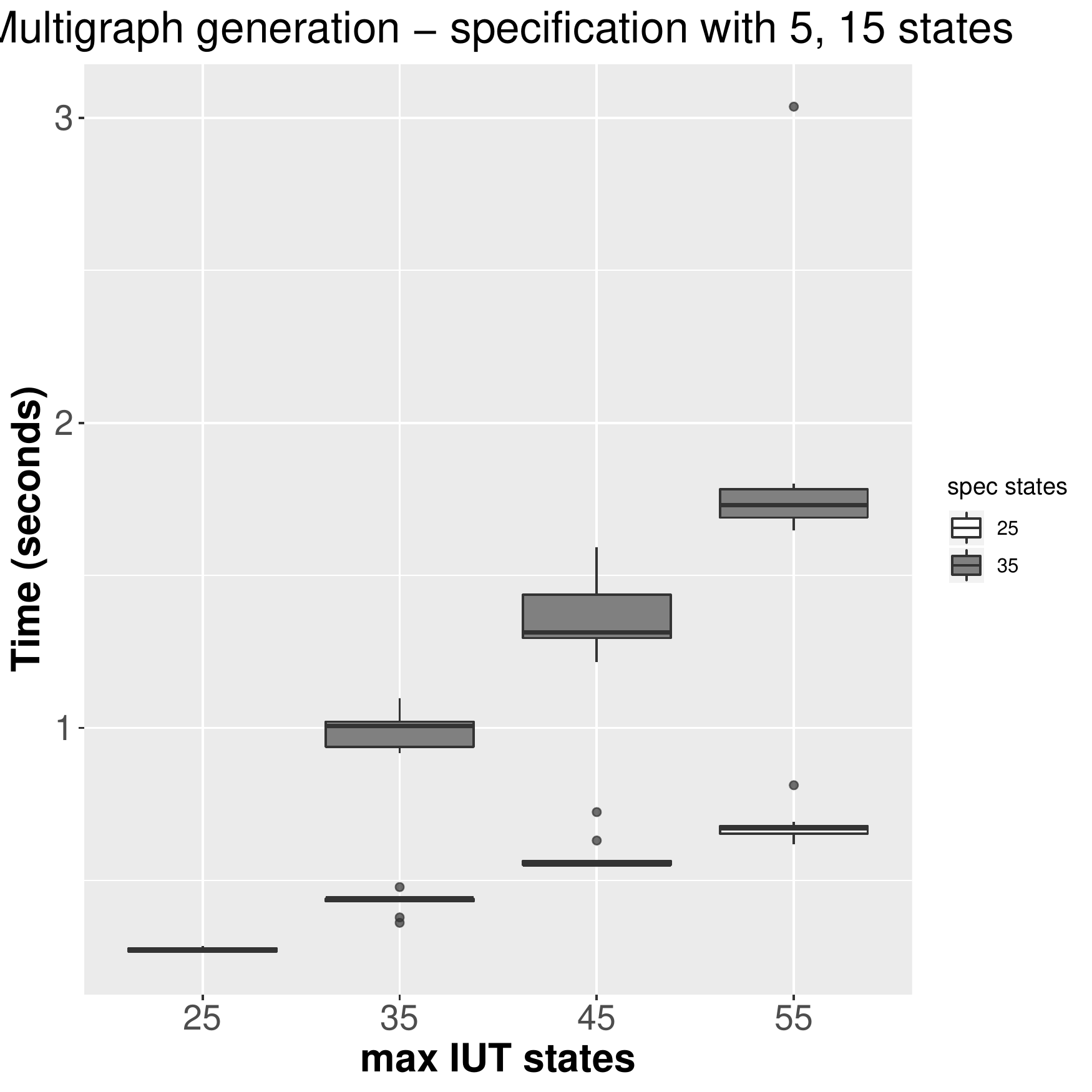}
			}\\						
		\end{tabular}
	\end{center}
	%	\vspace*{-2ex}
	\caption{Multigraph generation \label{fig:experimento-multigrafo} }
	%	\vspace*{-3ex}
\end{figure}

Figure~\ref{fig:experimento-multigrafo} shows that the processing time on generating multigraphs, in general, grows as the number of states also grows on specification and IUT models. 
%\begin{figure}[hbt]
%%		\vspace*{-2ex}
%	\begin{center}
%		\begin{tabular}{@{}cc@{}}
%			\subfloat[][Specifications with 5 and 15 states \label{fig:ger-mult-5-15}]{
%				\includegraphics[scale=0.3]{figs/experimentos/geracao/desempenho-ger-multigrafo-spec5-15.pdf}	
%			}
%			
%			&
%			\subfloat[Specifications with 25 and 35 states \label{fig:ger-mult-25-35}]{
%				\includegraphics[scale=0.3]{figs/experimentos/geracao/desempenho-ger-multigrafo-spec25-35.pdf}
%			}\\						
%		\end{tabular}
%		
%		%		\begin{tabular}{@{}cc@{}}
%		%			\subfloat[][Multi-graph generation - specifications with 25 states \label{fig:ger-mult-25}]{
%		%				\includegraphics[scale=0.4]{figs/experimentos/geracao/desempenho-ger-multigrafo-spec25.pdf}	
%		%			}
%		%			
%		%			&
%		%			\subfloat[Multi-graphs generation - specifications with 35 states \label{fig:ger-mult-35}]{
%		%				\includegraphics[scale=0.4]{figs/experimentos/geracao/desempenho-ger-multigrafo-spec35.pdf}
%		%			}\\						
%		%		\end{tabular}
%		
%	\end{center}
%%	\vspace*{-2ex}
%	\caption{Multigraph generation. \label{fig:experimento-multigrafo} }
%%	\vspace*{-3ex}
%\end{figure}
We can see at Figure~\ref{fig:ger-mult-5-15} that the multigraph construction for specifications with 5 states and $m=35$ takes 0.038 seconds on average, whereas the construction with $m=55$ takes 0.047 seconds. 
The processing time rose by 23,68$\%$. 
Similarly, we observe at Figure~\ref{fig:ger-mult-5-15} that the construction process for specifications with 15 states takes 0.186 seconds, on average, with $m=35$, and takes 0.253 seconds with $m=55$. 
In this case, the processing time rose by  36.02$\%$. 
Taking specifications with 25 states and $m=35$ the average time consumption of the multigraph construction process  is 0.428 seconds, and for $m=55$ it takes 0.676 seconds, as we can see in Figure~\ref{fig:ger-mult-25-35}. 
Now the processing time rose by 57.94$\%$. 
In the last group, we take specifications with 35 states and $m=35$, resulting a time of 0.994 seconds, in average, while it takes 1.867 seconds with $m=55$. 
Here the processing time rose by  87.82$\%$. 

Notice that the multigraph generation using the parameter $m=35$ is 46 times faster, on average, for specifications with  5 states than specifications with 35 states. 
Likewise the construction with $m=55$ is about 26 times faster for specifications with 5 states than specifications with 55 states.
We can conclude that the performance of the multigraph generation decreases as the number of states on specifications and the parameter $m$ increase. 
But the most important the processing time is not  meaningly affected, that is, the processing time does not substantially increase as we rise the number of states.

\subsubsection{TP Generation process}

Now we turn into the TP generation step based on multigraphs. 
So the associated \textbf{RQ} is: 
\textit{``How the TP generation is impacted w.r.t. the processing time when taking multigraphs where  the number of states of the corresponding specifications and the number of states to be considered over IUT models were varied?''}. 
In this case we take into account those multigraphs that were generated in the previous subsection using specifications with 5 to 35 states, and the parameter $m$ from 5 to 55. 
We fixed the number of TPs to be generated at 1000. 

\begin{figure}[hbt]
	%\vspace*{-2ex}
	\begin{center}
		\begin{tabular}{@{}cc@{}}
			\subfloat[][Specifications with 5 and 15 states \label{fig:ger-tp-5-15}]{
				\includegraphics[scale=0.4]{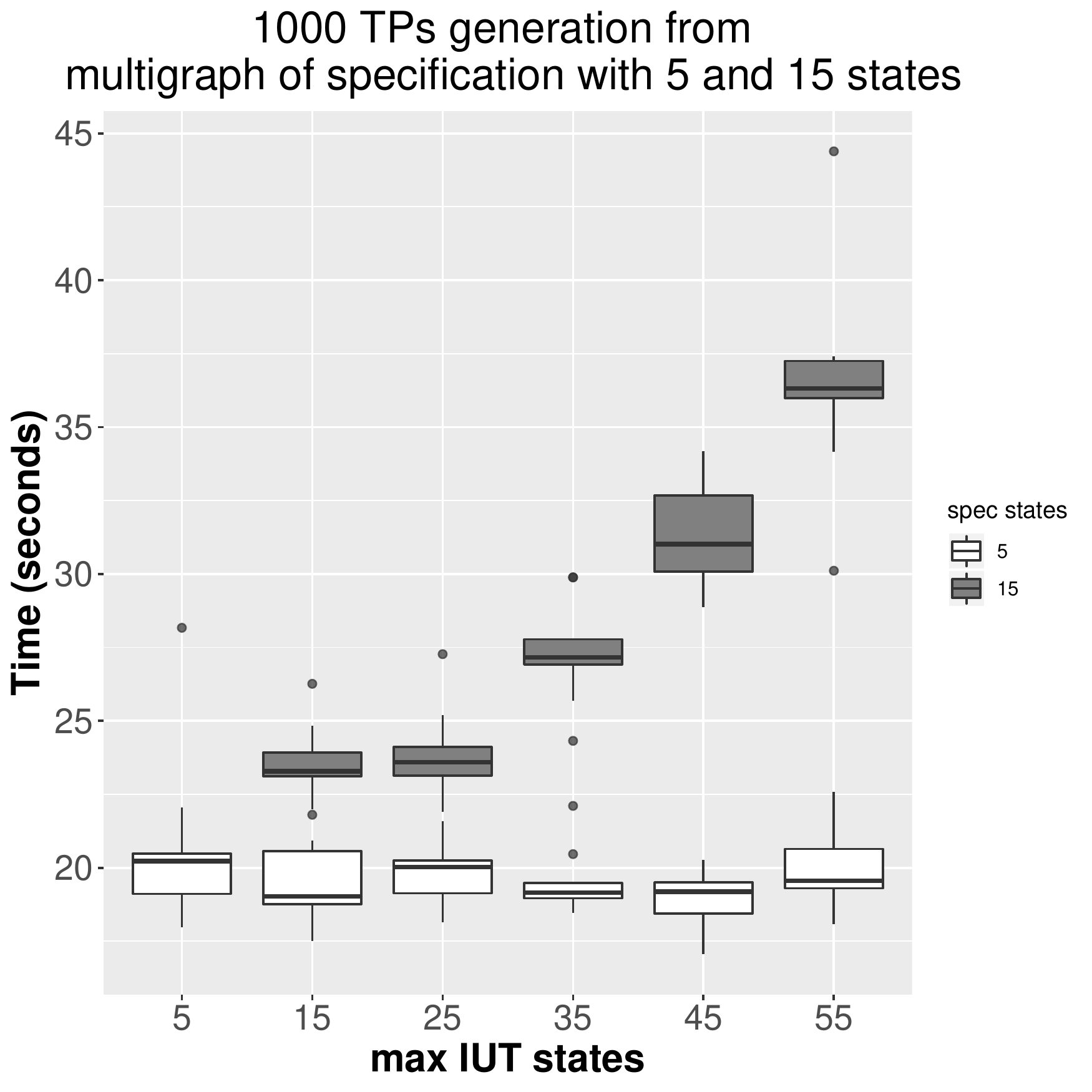}	
			}&
			\subfloat[Specifications with 25 and 35 states \label{fig:ger-tp-25-35}]{
				\includegraphics[scale=0.4]{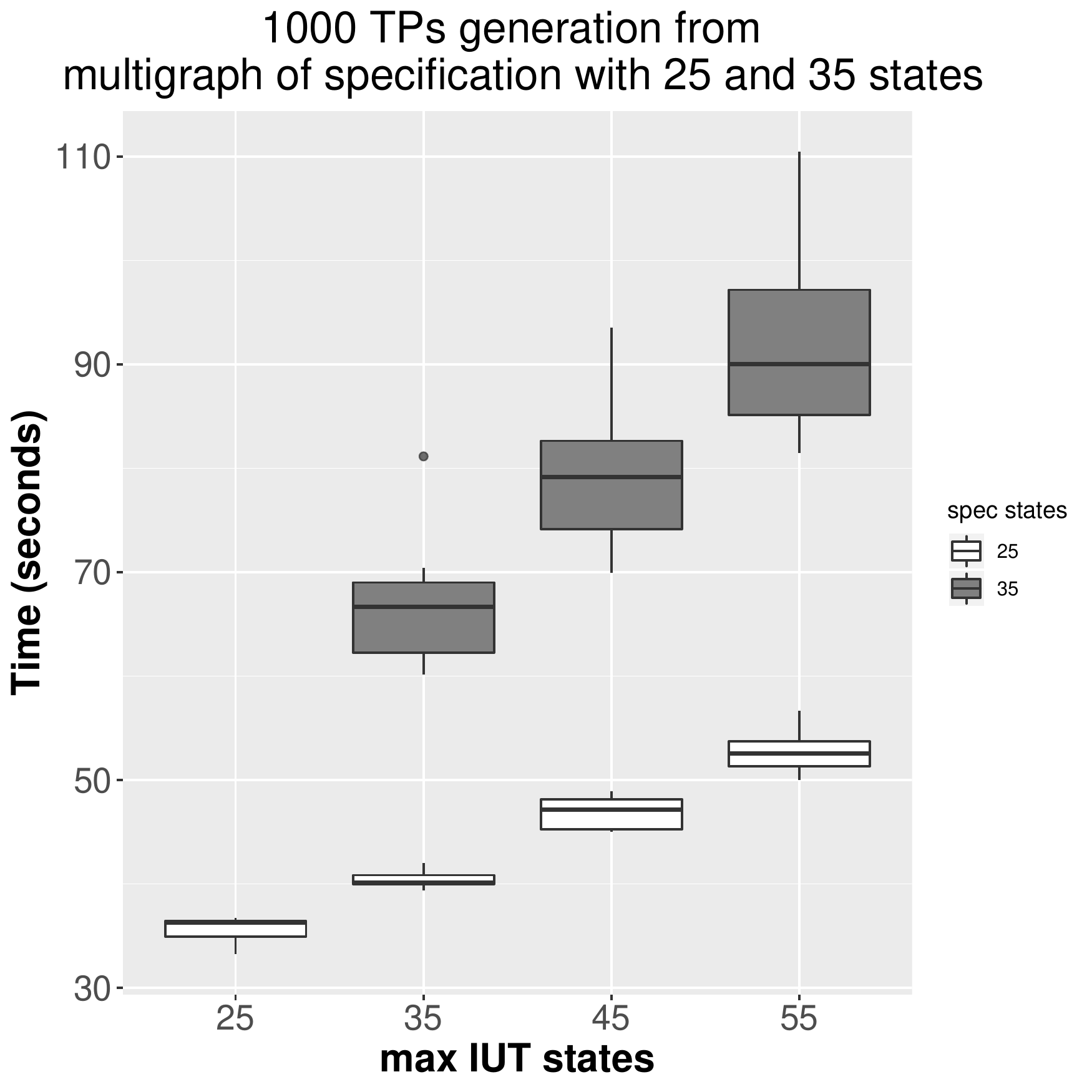}
			}\\						
		\end{tabular}		
	\end{center}
	%\vspace*{-2ex}
	\caption{TP generation \label{fig:experimento-tp} }
	%\vspace*{-3ex}
\end{figure}

Figure~\ref{fig:experimento-tp} shows that the test generation process takes much more time compared to the multigraph generation step. 
%\begin{figure}[hbt]
%		%\vspace*{-2ex}
%	\begin{center}
%		\begin{tabular}{@{}cc@{}}
%			\subfloat[][Specifications with 5 and 15 states \label{fig:ger-tp-5-15}]{
%				\includegraphics[scale=0.3]{figs/experimentos/geracao/desempenho-ger-tps-spec5-15.pdf}	
%			}
%			
%			&
%			\subfloat[Specifications with 25 and 35 states \label{fig:ger-tp-25-35}]{
%				\includegraphics[scale=0.3]{figs/experimentos/geracao/desempenho-ger-tps-spec25-35.pdf}
%			}\\						
%		\end{tabular}
%		
%		%		\begin{tabular}{@{}cc@{}}
%		%			\subfloat[][TP generation - specifications with 25 states \label{fig:ger-tp-25}]{
%		%				\includegraphics[scale=0.4]{figs/experimentos/geracao/desempenho-ger-tps-spec25.pdf}	
%		%			}
%		%			
%		%			&
%		%			\subfloat[TP generation - specifications with 35 states \label{fig:ger-tp-35}]{
%		%				\includegraphics[scale=0.4]{figs/experimentos/geracao/desempenho-ger-tps-spec35.pdf}
%		%			}\\						
%		%		\end{tabular}
%		
%	\end{center}
%	%\vspace*{-2ex}
%	\caption{TP generation. \label{fig:experimento-tp} }
%	%\vspace*{-3ex}
%\end{figure}
We also see at Figure~\ref{fig:ger-tp-5-15} that the processing time is more uniform for specifications with 5 states no matter we vary  $m$. 
When the number of states grows, the processing time of the TP generation grows faster as the parameter $m$ is increased, as we can see at Figure~\ref{fig:ger-tp-25-35}. 
The processing time for specifications with 35 states and $m=35$ takes 66.82 seconds whereas using $m=55$ it takes 91.67 seconds.
So we see that the rate rose by $37.19\%$. 
Considering specifications with 15 states and, respectively, with $m=35$ and $m=55$, the rate rose by 33.75$\%$, whereas for  specifications with 25 states and the same $m$ values the rate rose by 30.48$\%$.

\subsubsection{Running test suites}

In the last group of experiments we evaluate the time on running tests. 
Here the \textbf{RQ} is given as follows: 
\textit{``What is the impact on the processing time when running test suites over IUTs with 1\%, 2\% and  4\% of modification w.r.t. the specifications which were used to generate the respective multigraphs?''}. 
To answer this question we have taken test suites of TPs that were generated from specifications with 15 and 25 states. 
We fixed $m=n$, that is, the same number of states of the specifications.

\begin{figure}[hbt]
%		\vspace*{-1ex}
	\begin{center}
		\includegraphics[scale=0.5]{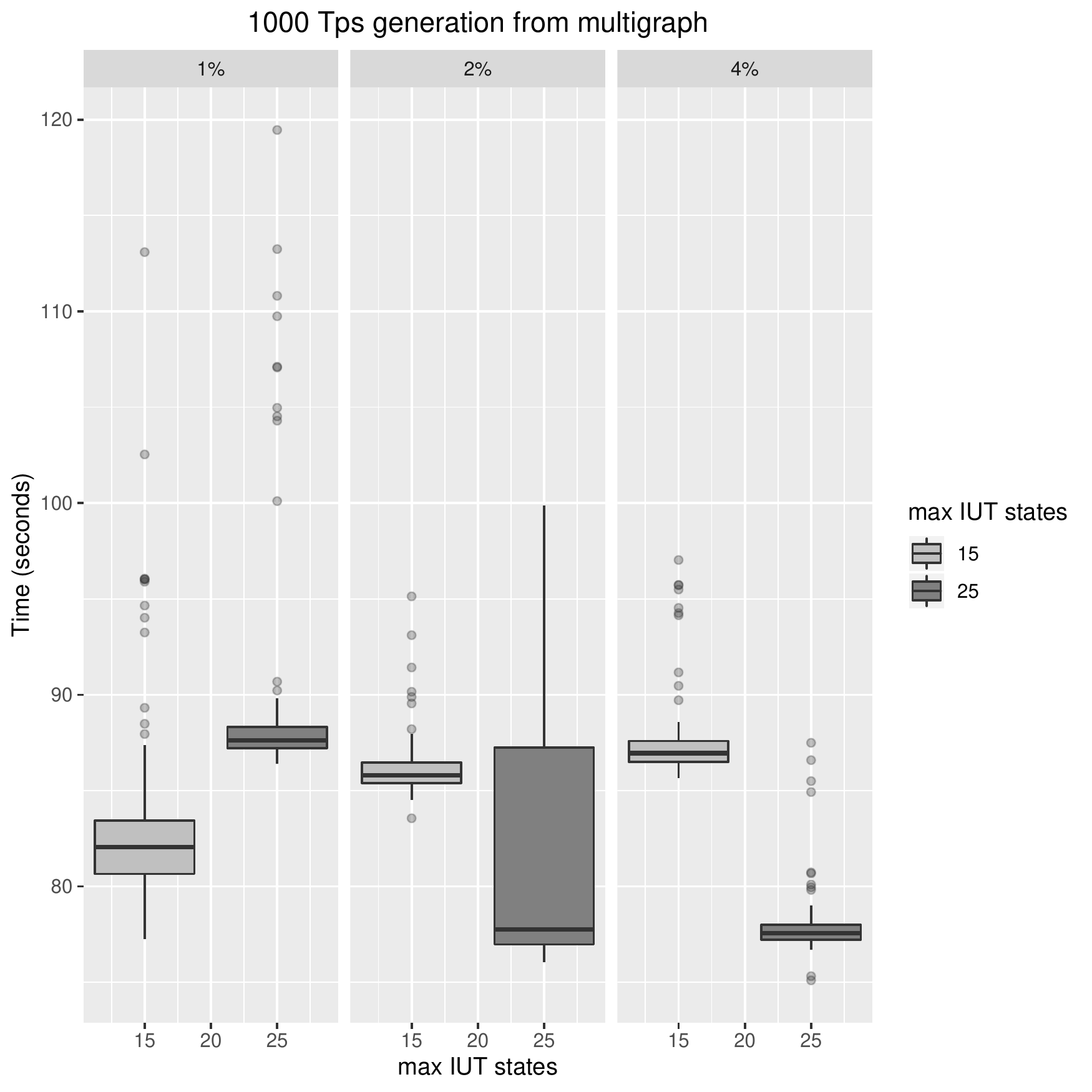}				
	\end{center}
%	\vspace*{-2ex}
	\caption{Test run \label{fig:experimento-run} }
%	\vspace*{-3ex}
\end{figure}

Figure~\ref{fig:experimento-run} shows the processing time according to the modification rate of IUTs. 
The time of the test run over IUTs with 1$\%$ of modification takes 83.03 seconds with $m=15$ and 89.78 seconds with $m=25$. 
Regarding IUTs with 2$\%$ of modification, the process takes 86.19 seconds with $m=15$ and 80.83 seconds with $m=25$. 
Finally, for IUTs with 4$\%$ of modification, the test run takes 87.54 seconds with $m=15$ and 77.83 seconds with $m=25$. 

We see that the time average of test runs over IUTs with 1$\%$ of modification is 5.15$\%$ faster than the test run over IUTs with 4$\%$ of modification setting $m=15$. 
If we consider $m=25$, the test run over IUTs with 4$\%$ of modification is 13.31$\%$ faster than over IUTs with 1$\%$ of modification.

%%%%%%%%%%%%%%%%%
\subsection{Threats to Validity}\label{subsec:threats}

We have listed some aspects that may arise as a threat to the validity of the experiments.  
We first relate a substantial difficulty when obtaining the JTorx source code to set both tools under the same conditions. 
% on performing the experiments. 
It would allow us to get more easily and precisely the time consumption on running the experiments. 
The computational resource that were used to run all experiments may also be a threat. 
We have used a general purpose machine which could have biased the results. 
But we remark that we have run both tools under theses same conditions. 

Another threat is related to the random generation of transitions over the models. 
Although we have randomly generated all models to keep the whole process unbiased, we must guarantee some properties on specific classes of experiments. 
For instance, we must construct IUTs that conform to the respective specifications in some groups of experiments, or we need to guarantee any modification over the IUTs to get verdicts of non-conformance. 
So these extra checking tasks somehow may bias the results. 

We also list as a threat properties that must guaranteed over the models following those restrictions imposed by JTorx tool. 
We see that the size of alphabets, the number of states and transitions of the specification and IUT models are modified from the original models in order to secure these properties. 
So we cannot make any claim about the similarity between these models and the original ones. 

%To overcome this threat, in future works, the experiments could be replicated using real-world FSMs as subjects. 
%
%To tackle this issue, we used a high number of different FSMs for each configuration, reducing the influence of this factor on the results.

% !TeX spellcheck = en_US
%\vspace*{-2ex}
\section{Conclusion}\label{sub:conclusao}

Conformance checking and test suite generation are important activities to improve the reliability  on developing reactive systems~\cite{belinfante1999}.
%Model-based testing has been widely used in the testing of reactive systems~\cite{aichernig2015,anand2013}, in particular, there are several works focused on the conformance checking~\cite{bonifacio2018,simao2014,tretmans2008} and test generation in order to increase the reliability of these systems.
In this work we have presented an automatic testing tool  for 
checking conformance and generating test suites for IOLTS models. 

We have implemented the classical \ioco relation and the more general approach based on regular languages. 
The latter, and consequently \everest tool, imposes few, if any, restrictions over the models and allows a wider range of fault models described by regular languages when checking conformance. 
Several works in the literature~\cite{calame2005,utting2007,belinfante2014,bhateja2014} have dealt with \ioco theory and its variations. 
However,  we are not aware of any other tool that implements a different notion of conformance, such as the language-based conformance. 
Here we have implemented a black-box test suite generation using the notion of test purposes. 

We described some case studies in order to probe both tools and their functionalities in practice. 
We then could observe from a comparative analysis that \everest provides a wider range of testing scenarios since it was able to detect faults, using the language-based approach, that were not detected by JTorx, using the \ioco theory.
The effectiveness  of our test suite generation method is also evaluated in the black-box scenarios.

We also offered practical experiments of conformance checking to compare the performance of JTorx and Everest.  
We can see that Everest outperforms JTorx in the most scenarios unless for those ones where the structure of IUT models are quite different from the corresponding specifications.
So we remark that although \everest implements a more general conformance relation the time consumption has not been impacted on 
checking runs. 
Also we observed from the results that \everest has a more stable behavior w.r.t. the processing time even for IUT models with quite different number of states. 
We also performed experiments of test suite generation and test run using \everest tool.  
%We note that our tool supports a test suite generation for specifications with up to 35 states and cover IUTs with up to 55 states. 
%That is, o
Our tool was able to handle specifications and implementation candidates with a reasonable number of states as seen in the experiments. 

We remark that our main contribution here is our practical tool that can check conformance  based on different relations and can generate test suites in a black-box setting. 
Moreover, we have presented some case studies, a comparative analysis, and also practical experiments to evaluate and compare our tool.

%%%%%%%%%%%%%%%%

%Como trabalhos futuros pretende-se investigar melhor alguns 
%comportamentos da ferramenta apresentados pelos gráficos dos experimentos, como por exemplo aquele apresentado no gráfico da Figura~\ref{fig:io-ioco-conf-2}, variando o número de entradas e saídas. 
%Neste caso, o tempo de verificação de conformidade da \everest diminui e na JTorx aumenta para IUTs a partir de 25 estados. 

As future work our tool's interface will be extended with a new module to allow conformance checking, test suite generation and test runs in a batch mode, \emph{i.e.}, we will be able to automatically test several IUT models. 
We also intend to improve our strategies and algorithms that generate test suites and also run test cases. 
The aim is to improve the \everest\!'s scalability and allow it to deal with larger models more efficiently. 

%Espera-se também que em extensões futuras a ferramenta possa gerar e testar IUTs caixa-preta de maneira direta através de protocolos de comunicação com sistemas reais. 

\end{document}